\documentclass[twocolumn, twocolappendix,
]{aastex631}

\usepackage{multirow}
\usepackage{soul}
\usepackage[capitalise]{cleveref}
\crefname{section}{Section}{Sections}
\Crefname{section}{Section}{Sections}
\crefname{equation}{Equation}{Equations}
\crefname{figure}{Figure}{Figures}

\makeatletter
\usepackage{etoolbox}
\patchcmd\H@refstepcounter{\protected@edef}{\protected@xdef}{}{}
\makeatother

\defcitealias{PB20}{PB20}
\newcommand{\PBXX}{\citetalias{PB20}}

\shorttitle{Improved measurement of degree-scale CMB $B$-mode polarization with Polarbear}
\shortauthors{Polarbear Collaboration}
\graphicspath{{./}{figures/}}

\begin{document}

\title{Improved upper limit on degree-scale CMB $B$-mode polarization power from the 670 square-degree P\footnotesize{}OLARBEAR\normalsize{} survey}

\correspondingauthor{S.~Takakura}
\email{satoru.takakura-1@colorado.edu}

\collaboration{0}{The \textsc{Polarbear} collaboration}

\author[0000-0002-0400-7555]{S.~Adachi}
\affiliation{Department of Physics, Faculty of Science, Kyoto University, Kyoto, 606-8502, Japan}
\affiliation{Hakubi Center for Advanced Research, Kyoto University, Yoshida Honmachi, Sakyo-ku, Kyoto, 606-8501, Japan}
\author[0000-0002-3850-9553]{T.~Adkins}
\affiliation{Department of Physics, University of California, Berkeley, Berkeley, CA 94720, USA}
\author[0000-0002-1571-663X]{M.~A.~O.~Aguilar Faúndez}
\affiliation{Departamento de Evaluación, Medición y Registro Educacional, Universidad de Chile, Santiago, 7750358, Chile}
\author[0000-0002-3407-5305]{K.~S.~Arnold}
\affiliation{Department of Physics, University of California, San Diego, La Jolla, CA 92093-0424, USA}
\author[0000-0002-8211-1630]{C.~Baccigalupi}
\affiliation{Astrophysics and Cosmology, International School for Advanced Studies (SISSA), Trieste, I-34136, Italy}
\affiliation{Institute for Fundamental Physics of the Universe (IFPU), Via Beirut 2, 34014 Trieste, Italy}
\affiliation{National Institute for Nuclear Physics (INFN), Sezione di Trieste, Padriciano, 99, I-34149 Trieste, Italy}
\author[0000-0002-1623-5651]{D.~Barron}
\affiliation{Department of Physics and Astronomy, University of New Mexico, Albuquerque, NM 87131, USA}
\author[0000-0002-8487-3153]{S.~Chapman}
\affiliation{Physics and Atmospheric Science, Dalhousie University, Halifax, B3H4R2, Canada}
\affiliation{Herzberg Astronomy and Astrophysics, National Research Council, Victoria, V9E 2E7, Canada}
\author[0000-0002-7764-378X]{K.~Cheung}
\affiliation{Department of Physics, University of California, Berkeley, Berkeley, CA 94720, USA}
\affiliation{Space Sciences Laboratory, University of California, Berkeley, Berkeley, CA 94720, USA}
\affiliation{Computational Cosmology Center, Lawrence Berkeley National Laboratory, Berkeley, CA 94720, USA}
\author[0000-0002-3266-857X]{Y.~Chinone}
\affiliation{Research Center for the Early Universe, Graduate School of Science, The University of Tokyo, Tokyo, 113-0033, Japan}
\affiliation{Kavli Institute for the Physics and Mathematics of the Universe (WPI), UTIAS, The University of Tokyo, Kashiwa, Chiba, 277-8583, Japan}
\author[0000-0001-5068-1295]{K.~T.~Crowley}
\affiliation{Department of Physics, University of California, Berkeley, Berkeley, CA 94720, USA}
\author[0000-0002-5166-5614]{T.~Elleflot}
\affiliation{Physics Division, Lawrence Berkeley National Laboratory, Berkeley, CA 94720, USA}
\author[0000-0002-1419-0031]{J.~Errard}
\affiliation{Universit'{e} de Paris, CNRS, Astroparticule et Cosmologie, F-75013 Paris, France}
\author[0000-0002-3255-4695]{G.~Fabbian}
\affiliation{School of Physics and Astronomy, Cardiff University, Cardiff, The Parade, CF24 3AA, UK}
\affiliation{Center for Computational Astrophysics, Flatiron Institute, 162 5th Avenue, NY 10010, USA}
\author[0000-0001-7438-5896]{C.~Feng}
\affiliation{School of Astronomy and Space Science, University of Science and Technology of China, HeFei, 230026, People's Republic of China}
\author[0000-0002-1211-7850]{T.~Fujino}
\affiliation{Graduate School of Engineering Science, Yokohama National University, Yokohama, 240-8501, Japan}
\author[0000-0001-7225-6679]{N.~Galitzki}
\affiliation{Department of Physics, University of California, San Diego, La Jolla, CA 92093-0424, USA}
\author[0000-0003-2606-9340]{N.~W.~Halverson}
\affiliation{CASA \& Department of Astrophysical \& Planetary Sciences, University of Colorado Boulder, Boulder, CO 80309, USA}
\author[0000-0003-1443-1082]{M.~Hasegawa}
\affiliation{Institute of Particle and Nuclear Studies, High Energy Accelerator Research Organization (KEK), Tsukuba, 305-0801, Japan}
\author[0000-0001-6830-8309]{M.~Hazumi}
\affiliation{International Center for Quantum-field Measurement Systems for Studies of the Universe and Particles (QUP), High Energy Accelerator Research Organization (KEK), Tsukuba, 305-0801, Japan}
\affiliation{Institute of Particle and Nuclear Studies, High Energy Accelerator Research Organization (KEK), Tsukuba, 305-0801, Japan}
\affiliation{Institute of Space and Astronautical Science, Japan Aerospace Exploration Agency (JAXA), Sagamihara, 252-5210, Japan}
\affiliation{Kavli Institute for the Physics and Mathematics of the Universe (WPI), UTIAS, The University of Tokyo, Kashiwa, Chiba, 277-8583, Japan}
\affiliation{School of High Energy Accelerator Science, The Graduate University for Advanced Studies, SOKENDAI, Kanagawa, 240-0193, Japan}
\author{H.~Hirose}
\affiliation{Graduate School of Engineering Science, Yokohama National University, Yokohama, 240-8501, Japan}
\author[0000-0003-1887-5860]{L.~Howe}
\affiliation{Superconductive Electronics Group, National Institute of Standards and TEchnology, Boulder, 80305, USA}
\affiliation{Department of Physics, University of California, San Diego, La Jolla, CA 92093-0424, USA}
\author{J.~Ito}
\affiliation{Department of Physics, University of California, San Diego, La Jolla, CA 92093-0424, USA}
\author[0000-0001-5893-7697]{O.~Jeong}
\affiliation{Department of Physics, University of California, Berkeley, Berkeley, CA 94720, USA}
\author[0000-0003-3917-086X]{D.~Kaneko}
\affiliation{Institute of Particle and Nuclear Studies, High Energy Accelerator Research Organization (KEK), Tsukuba, 305-0801, Japan}
\author{N.~Katayama}
\affiliation{Kavli Institute for the Physics and Mathematics of the Universe (WPI), UTIAS, The University of Tokyo, Kashiwa, Chiba, 277-8583, Japan}
\author[0000-0003-3118-5514]{B.~Keating}
\affiliation{Department of Physics, University of California, San Diego, La Jolla, CA 92093-0424, USA}
\author[0000-0003-3510-7134]{T.~Kisner}
\affiliation{Computational Cosmology Center, Lawrence Berkeley National Laboratory, Berkeley, CA 94720, USA}
\affiliation{Department of Physics, University of California, Berkeley, Berkeley, CA 94720, USA}
\author{N.~Krachmalnicoff}
\affiliation{Astrophysics and Cosmology, International School for Advanced Studies (SISSA), Trieste, I-34136, Italy}
\affiliation{Institute for Fundamental Physics of the Universe (IFPU), Via Beirut 2, 34014 Trieste, Italy}
\affiliation{National Institute for Nuclear Physics (INFN), Sezione di Trieste, Padriciano, 99, I-34149 Trieste, Italy}
\author{A.~Kusaka}
\affiliation{Physics Division, Lawrence Berkeley National Laboratory, Berkeley, CA 94720, USA}
\affiliation{Department of Physics, The University of Tokyo, Tokyo, 113-0033, Japan}
\affiliation{Kavli Institute for the Physics and Mathematics of the Universe (WPI), Berkeley Satellite, University of California, Berkeley, Berkeley, CA 94720, USA}
\affiliation{Research Center for the Early Universe, Graduate School of Science, The University of Tokyo, Tokyo, 113-0033, Japan}
\author[0000-0003-3106-3218]{A.~T.~Lee}
\affiliation{Department of Physics, University of California, Berkeley, Berkeley, CA 94720, USA}
\affiliation{Physics Division, Lawrence Berkeley National Laboratory, Berkeley, CA 94720, USA}
\affiliation{Radio Astronomy Laboratory, University of California, Berkeley, Berkeley, CA 94720, USA}
\author[0000-0001-5536-9241]{E.~Linder}
\affiliation{Space Sciences Laboratory, University of California, Berkeley, Berkeley, CA 94720, USA}
\affiliation{Physics Division, Lawrence Berkeley National Laboratory, Berkeley, CA 94720, USA}
\author[0000-0003-1200-9179]{A.~I.~Lonappan}
\affiliation{Astrophysics and Cosmology, International School for Advanced Studies (SISSA), Trieste, I-34136, Italy}
\affiliation{Institute for Fundamental Physics of the Universe (IFPU), Via Beirut 2, 34014 Trieste, Italy}
\affiliation{National Institute for Nuclear Physics (INFN), Sezione di Trieste, Padriciano, 99, I-34149 Trieste, Italy}
\author{L.~N.~Lowry}
\affiliation{Department of Physics, University of California, Berkeley, Berkeley, CA 94720, USA}
\author[0000-0003-0041-6447]{F.~Matsuda}
\affiliation{Institute of Space and Astronautical Science, Japan Aerospace Exploration Agency (JAXA), Sagamihara, 252-5210, Japan}
\author[0000-0001-9002-0686]{T.~Matsumura}
\affiliation{Kavli Institute for the Physics and Mathematics of the Universe (WPI), UTIAS, The University of Tokyo, Kashiwa, Chiba, 277-8583, Japan}
\author[0000-0003-2176-8089]{Y.~Minami}
\affiliation{Research Center for Nuclear Physics, Osaka University, Ibaraki, Osaka, 567-0047, Japan}
\author[0000-0003-4394-4645]{M.~Murata}
\affiliation{Department of Physics, The University of Tokyo, Tokyo, 113-0033, Japan}
\author[0000-0003-0738-3369]{H.~Nishino}
\affiliation{Information-Technology Promotion Division, Japan Synchrotron Radiation Research Institute (JASRI), Sayo-cho, 679-5198, Japan}
\author[0000-0001-9204-3611]{Y.~Nishinomiya}
\affiliation{Department of Physics, The University of Tokyo, Tokyo, 113-0033, Japan}
\author[0000-0001-9807-3758]{D.~Poletti}
\affiliation{Department of Physics ``Giuseppe Occhialini,'' Milano-Bicocca University, Milan, I-20126, Italy}
\author[0000-0003-2226-9169]{C.~L.~Reichardt}
\affiliation{School of Physics, The University of Melbourne, Parkville, VIC, 3010, Australia}
\author{C.~Ross}
\affiliation{Physics and Atmospheric Science, Dalhousie University, Halifax, B3H4R2, Canada}
\author{Y.~Segawa}
\affiliation{School of High Energy Accelerator Science, The Graduate University for Advanced Studies, SOKENDAI, Kanagawa, 240-0193, Japan}
\author[0000-0001-6830-1537]{P.~Siritanasak}
\affiliation{National Astronomical Research Institute of Thailand, Chiangmai, 50180, Thailand}
\author[0000-0002-9777-3813]{R.~Stompor}
\affiliation{CNRS-UCB, International Research Laboratory, Centre Pierre Binétruy, IRL2007, CPB-IN2P3, Berkeley, CA 94720, USA}
\affiliation{Universit'{e} de Paris, CNRS, Astroparticule et Cosmologie, F-75013 Paris, France}
\author[0000-0001-8101-468X]{A.~Suzuki}
\affiliation{Physics Division, Lawrence Berkeley National Laboratory, Berkeley, CA 94720, USA}
\author{O.~Tajima}
\affiliation{Department of Physics, Faculty of Science, Kyoto University, Kyoto, 606-8502, Japan}
\author[0000-0001-9461-7519]{S.~Takakura}
\affiliation{CASA \& Department of Astrophysical \& Planetary Sciences, University of Colorado Boulder, Boulder, CO 80309, USA}
\author{S.~Takatori}
\affiliation{School of High Energy Accelerator Science, The Graduate University for Advanced Studies, SOKENDAI, Kanagawa, 240-0193, Japan}
\affiliation{Institute of Particle and Nuclear Studies, High Energy Accelerator Research Organization (KEK), Tsukuba, 305-0801, Japan}
\author{D.~Tanabe}
\affiliation{Center of High Energy and High Field Physics, National Central University, Taoyuan, 32001, Taiwan}
\affiliation{Institute of Particle and Nuclear Studies, High Energy Accelerator Research Organization (KEK), Tsukuba, 305-0801, Japan}
\author{G.~Teply}
\affiliation{Department of Physics, University of California, San Diego, La Jolla, CA 92093-0424, USA}
\author[0000-0003-0221-2130]{K.~Yamada}
\affiliation{Department of Physics, The University of Tokyo, Tokyo, 113-0033, Japan}
\author[0000-0002-5878-4237]{Y.~Zhou}
\affiliation{Department of Physics, University of California, Berkeley, Berkeley, CA 94720, USA}

\begin{abstract}

We report an improved measurement of the degree-scale cosmic microwave background $B$-mode angular-power spectrum over $670\,\mathrm{deg}^2$ sky area at $150\,\mathrm{GHz}$ with \textsc{Polarbear}. In the original analysis of the data, errors in the angle measurement of the continuously rotating half-wave plate, a polarization modulator, caused significant data loss. By introducing an angle-correction algorithm, the data volume is increased by a factor of $1.8$. We report a new analysis using the larger data set. We find the measured $B$-mode spectrum is consistent with the $\Lambda$CDM model with Galactic dust foregrounds. 
We estimate the contamination of the foreground by cross-correlating our data and Planck 143, 217, and $353\,\mathrm{GHz}$ measurements, where its spectrum is modeled as a power law in angular scale and a modified blackbody in frequency.
We place an upper limit on the tensor-to-scalar ratio $r < 0.33$ at 95\% confidence level after marginalizing over the foreground parameters.

\end{abstract}


\section{Introduction} \label{sec:intro}
Anisotropies in the cosmic microwave background (CMB) bring us fundamental information about our universe. 
If detected, degree-scale $B$-mode polarization, the parity-odd component of the linear polarization anisotropies, is a footprint of the primordial gravitational waves generated during the cosmic inflation era. By measuring the amplitude of the $B$-modes, we can determine the tensor-to-scalar ratio $r$ and test the physical mechanisms of the inflation.

Current 95\% upper limits on $r$ are 0.036 from \citet{BK2021}, 0.044 from \citet{Tristram2021}, 
0.11 from \citet{SPIDER2021}, 0.44 from \citet{SPTpol2020}, 0.90 from \citet{PB20}, and 2.3 from ABS~\citep{ABS2018}.

The \textsc{Polarbear} experiment is a ground-based experiment in the Atacama desert in Chile. It consists of the $2.5\,\mathrm{m}$ aperture Huan Tran Telescope with 1274 transition-edge-sensor bolometers sensitive to the $150\,\mathrm{GHz}$ band \citep{Ziggy2012SPIE,Kam2012SPIE}. In 2014, we installed a continuously rotating half-wave plate (HWP) at the prime focus~\citep{Takakura2017JCAP}. 
The HWP modulates incoming linear polarization signals and therefore reduces low-frequency noise due to both the atmosphere and the instrument.

In \citet[][hereafter \PBXX{}]{PB20}, we reported a measurement of the degree-scale $B$-mode angular-power spectrum using data from three years of observations from 2014 to 2016.
The HWP modulation results in a relatively low knee in the noise spectrum at $\ell_\mathrm{knee} = 90$, where the contribution of the low-frequency noise to the power spectrum uncertainty becomes comparable to that of detector white noise.
We place an upper limit of $r<0.90$ at the 95\% confidence level. 

In \PBXX{}, however, we used only 29.2\% of data after eliminating data from detectors that failed to tune correctly or that have glitches due to various disturbances.
Data containing glitches can potentially be made available by improvements to the analysis process.
We find that most of the glitches in the detector polarization timestream come from an angle error of the HWP.
Here, the angle error is the offset of the measured angle from the real angle, 
which occasionally occurs due to electrical noise within the encoder circuit.
By improving the encoder error correction, we successfully bypassed the glitches and recovered about 80\% more data. 

In this paper, we report an improved analysis with this revised data set.
We perform the same analysis pipeline as in \PBXX{}. 
Calibration of the pointing model and each detector's properties (pointing offset, relative gain, and relative polarization angle) is the same as in
\PBXX{}. 
Calibration using observed power spectra, i.e., absolute gain, beam smearing due to pointing jitter, and absolute polarization angle, 
is updated and found to be consistent with \PBXX{}. We confirm that the new data set passes the same set of null tests as
\PBXX{}. We assume that the systematic uncertainties are the same as \PBXX{} because we use the same seasons of data and 
calibration.
Note that statistical uncertainties are still dominant even with the additional data. Finally, we cross-correlate the \textsc{Polarbear} map with Planck 143, 217, and $353\,\mathrm{GHz}$ maps and estimate constraints on $r$ considering Galactic dust foregrounds as in
\PBXX{}.

In \cref{sec:glitch_dataselection} we explain the glitches due to the HWP angle error and improvements in the data processing.
The impact of the improved data processing on data selection is presented in \cref{sec:dataselection}.
In \cref{sec:pipeline} we follow the
\PBXX{} analysis pipeline and report absolute 
calibration,
null tests, and final power spectra. We perform parameter estimations in \cref{sec:pe}. In \cref{sec:comparison} we discuss consistency between \PBXX{} and the new results. Finally, we summarize in \cref{sec:conclusion}.

\section{Detector and HWP encoder data processing\label{sec:glitch_dataselection}}

The main improvement in this study is in the processing of the encoder data of the HWP angle.
In this section, we explain details of the HWP angle error:
how the angle error causes a glitch on the detector signal, 
how the angle error is caused, 
and how we have improved the correction of the angle error.

Except for the improved correction of the HWP encoder data, we follow the data processing presented in \PBXX{} and references therein. 

\subsection{Glitches Due to the Angle Error of the HWP \label{sec:glitch}} 
Glitches are spurious signals in detector timestreams. They have several causes:
transient physical events such as cosmic-ray hits, atmospheric noise in bad weather conditions, electrical noise pickup, and unexpected data drop in the readout system. Thus, glitches have various timescales and shapes. To drop all kinds of glitches, we apply several filters in our analysis process.

We apply glitch detection for three types of timestreams:
full-sampling timestreams for each detector, 
demodulated and downsampled timestreams for each detector, 
and timestreams averaged among all detectors.
In the first step, we detect short-timescale glitches such as cosmic ray hits. In the next step, we focus on glitches below $4\,\mathrm{Hz}$ after demodulation,\footnote{We modulate polarization signals at $8\,\mathrm{Hz}$ using the HWP. Thus, the glitches are in the range of 4--$12\,\mathrm{Hz}$, originally.} which contaminate our science signal.
Electrical pickup and bad weather data are flagged. Finally, we catch faint but correlated glitches. Polarized bursts due to clouds \citep{Takakura2019ApJ} are detected here. The common-mode glitch detection has the largest impact on the data selection because it has the highest sensitivity and affects all detectors.

The angle error of the HWP is another source of correlated glitches.
In observations with the rotating HWP, 
the detector signal, $d(t)$, 
is modeled as~\citep{Takakura2017JCAP}
\begin{equation}\label{eq:d(t)}
\begin{array}{rcl}
d(t) & = & I(t) + \mathrm{Re}[(Q(t) + i U(t))\,\mathrm{exp}(-4i\theta(t))] \\
& & + N(t) + \sum_{n} \mathrm{Re}[A_{n}(t)\,\mathrm{exp}(-i n \theta(t))]\;.
\end{array}
\end{equation}
The unpolarized Stokes component, $I(t)$, is not modulated, while the linear polarization components, $Q(t)$ and $U(t)$, are modulated by the angle of HWP, $\theta(t) = \omega t$, where $\omega/2\pi=2.0\,\mathrm{Hz}$.
$N(t)$ is detector noise. The last term is instrumental signals called HWP synchronous signals (HWPSSs), which are classified by the order of the harmonic $n$. We obtain polarization components by demodulating $d(t)$ as
\begin{equation}\label{eq:demod}
d_m(t) = F_\mathrm{LP}[2\,\mathrm{exp}(4i\theta'(t))\,F_\mathrm{BP}[d(t)]]\;,
\end{equation}
where $F_\mathrm{LP}$ and $F_\mathrm{BP}$ are low-pass and bandpass filters used to select signals around the modulation frequency $4\omega$. Here, we use the measured angle of the HWP $\theta'(t)$, 
which is reconstructed from the HWP encoder data. 
If the measured angle has an error from the actual angle as $\Delta\theta(t) = \theta'(t) - \theta(t)$,
the demodulated signal becomes
\begin{equation}\label{eq:dm(t)}
\begin{array}{rcl}
d_m(t) & \approx & Q(t) + i U(t) + N_Q(t) + iN_U(t) \\
& & + A_{4}(t) + 4i \langle A_{4}\rangle \Delta\theta(t)\;,
\end{array}
\end{equation}
where $N_Q(t)$ and $N_U(t)$ are demodulated detector noise, $A_{4}(t)$ is the fourth harmonic of the HWPSS, and $\langle A_{4}\rangle$ is its average.
Here, we assume that $\Delta\theta(t) \ll 1$ and $\langle A_{4}\rangle$ is much larger than $Q(t)$ and $U(t)$.
In the case of \textsc{Polarbear}, instrumental polarization due to the primary mirror produces $A_4 \sim 0.1\,\mathrm{K}$ uniformly among all detectors~\citep{Takakura2017JCAP}. Therefore, the last term of \cref{eq:dm(t)} becomes a source of correlated glitches.

\subsection{Example of Data with the Angle Error\label{sec:glitch_example}}
\begin{figure}[t!]
\epsscale{1.15}
\plotone{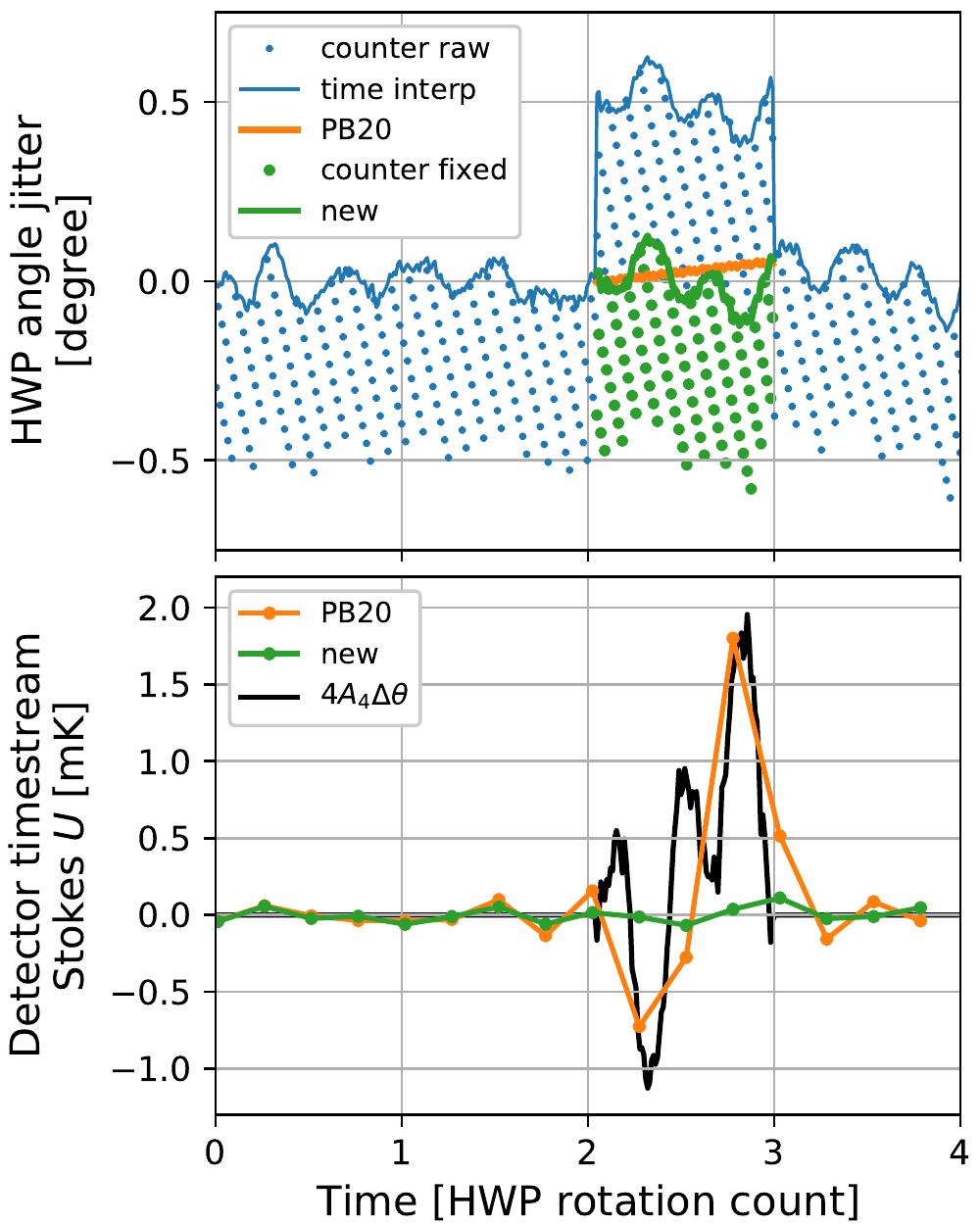}
\caption{
Example of encoder data with a wrong count and corresponding detector signal. The horizontal axis is the rotation count of the HWP, which takes $0.5\,\mathrm{s}$ per rotation. The top panel shows HWP encoder data subtracted by the regular drift, $\omega t$. 
The blue points are raw encoder counts sampled at $191\,\mathrm{Hz}$. 
The uniform pattern is quantization noise due to the resolution of the encoder plate by $0\fdg5$. 
The blue line shows the interpolated angle using a timestamp.
The wavy pattern on it is the actual rotation jitter of the HWP. 
In the third rotation, it jumps by $0\fdg5$ due to an error count and is reset by the indexing signal. 
The orange line shows the correction in \PBXX{}, which linearly interpolates the jump.
The green points show the new method, in which we fix the error at the encoder count level and then apply timestamp interpolation.
The green line is the result, which keeps the actual HWP angle jitter.
The bottom panel shows the detector signal of Stokes $U$ component in the instrumental coordinates. 
\PBXX{} data, the orange line, show a glitch, which is consistent with the expectation by \cref{eq:dm(t)}, the black line. In the new data, the green line, the HWPSS is subtracted correctly, and the glitch disappears.
\label{fig:encodercorrection}}
\end{figure}

\Cref{fig:encodercorrection} shows an example of HWP encoder data causing the angle error $\Delta\theta(t)$ and 
the resulting glitch in the detector timestream. To understand this, 
it is necessary to explain how the encoder works.

We measure the angle of the HWP using an encoder plate with 360 precisely machined teeth and one index hole. Optocouplers regularly sample whether the gate is open or closed at $40\,\mathrm{kHz}$.
Synchronizing with detector sampling at $191\,\mathrm{Hz}$, 
we store the count of the edges and the timing of the last edge.
Since the timings of the edge and detector sampling are asynchronous, the raw encoder count (the blue points) has a quantization noise of $0\fdg5$.
We fix this quantization error by interpolating the middle angle at the detector sampling between edges using the timestamp information (the blue line).
The statistical uncertainty of the angle (the width of the blue line) comes from the quantization error on the timestamp of $3\times10^{-6}\,\mathrm{rad}\sqrt{\mathrm{s}}$.
This uncertainty causes an angle error noise of $1.4\,\mu\mathrm{K}\sqrt{\mathrm{s}}$, 
which is smaller than the detector array sensitivity.
On the other hand, the HWP rotation is not perfectly continuous and contains a jitter of $\delta\theta(t) = \theta(t) - \omega t$.
The wavy fluctuation of the blue line is the jitter, which has a weak resonance around $8\,\mathrm{Hz}$ probably due to the combination of the spring constant of the system and feedback parameters for the servo motor driver.
This actual jitter does not introduce noise on the demodulated signal if the encoder measures the jitter as $\theta'(t) = \theta(t) = \omega t + \delta\theta(t)$ and $\Delta\theta(t) = 0$.

The problem in our HWP encoder is that it occasionally has erroneous counts due to electrical noise (the jump in the blue line). We detect these bad counts by comparing two optocouplers. In \PBXX{}, we dropped encoder samples from the bad counts to the next index signal and interpolated linearly (the orange line). This interpolation nicely tracks the continuous rotation, but not the jitter, i.e., $\theta'(t) = \omega t$. The angle error in the interpolated samples becomes $\Delta\theta(t) = \delta\theta(t)$.
As shown in the bottom panel of \Cref{fig:encodercorrection}, this angle error causes the glitches (the orange points) by the last term of \cref{eq:dm(t)} (the black line).
Here, the amplitude of the glitch is $\sim 1\,\mathrm{mK}$, which is comparable to the instantaneous sensitivity of a single detector. This means that we cannot detect this glitch in a single detector analysis. However, this glitch is correlated among detectors, but the white noise is not. Therefore, averaging hundreds of detectors improves the significance dramatically. 

We improved the correction method to directly decrement the counter in the offline analysis (the green points).
Then, we apply quantization noise reduction as other normal data (the green line).
In this method, the fixed angle keeps information of the actual jitter as $\theta'(t) = \omega t + \delta\theta(t)$, and the angle error $\Delta\theta(t)$ becomes zero.
Therefore, we can clean the glitches as the green points in the bottom panel of \Cref{fig:encodercorrection}.

We reprocessed all data with the new method and successfully cleaned this type of glitch as described in \cref{sec:dataselection}.

To mitigate the risk of this type of error in HWP data acquisition in future experiments such as the Simons Array experiment~\citep{SA2016JLTP}, 
there are two solutions. One simple way is making the hardware for the encoder data acquisition robust.
In PB-2a, the first receiver of the Simons Array,
we use a commercial encoder~\citep{PB2aHWPSPIE}. The other solution is making the rotation stable, which makes the data more robust for offline data correction even if there were some faulty data.
The cryogenic superconducting bearing technique developed for PB-2b, the second receiver of the Simons Array,
is promising to achieve very stable rotation~\citep{PB2bHWP}. 

\section{Data selection\label{sec:dataselection}}
We applied the same \PBXX{} data selection method to the data processed with the new encoder correction. 
The results of data volume and data selection efficiencies are summarized in \cref{tab:datavolume,tab:dataselection}. 
The final volume of data available increased by 81.6\% from \PBXX{}.
\begin{deluxetable}{lrrr}
\tabletypesize{\footnotesize}
\tablecaption{Data Volume\label{tab:datavolume}}
\tablehead{ & & & \colhead{\scriptsize Fractional}\\[-10pt]
& \colhead{New} & \colhead{PB20} & \colhead{\scriptsize Change}}
\startdata
Observation & & & \\
\hspace{10pt}from & 2014 Jul 25 & 2014 Jul 25 & \\
\hspace{10pt}until & 2016 Dec 30 & 2016 Dec 6 & \\
Total calendar time & 21,340 hr & 20,766 hr & $+2.8$\% \\
Time observing patch & 6818 hr & 6610 hr & $+3.1$\% \\ 
Observation efficiency & 31.9\% & 31.8\% & \\\hline Total detector & 1274 & 1274 & $0$\% \\
With calibration & 647 & 647 & $0$\% \\
Detector yield & 50.8\% & 50.8\% & \\\hline Total volume of data & 4,410,986 hr & 4,276,467 hr & $+3.1$\% \\
Final volume of data & 1,597,098 hr & 879,235 hr & $+81.6$\% \\
Data selection efficiency\hspace*{-20pt} & 36.2\% & 20.6\% & \\\hline\hline Overall efficiency & 5.87\% & 3.32\% & \\
\enddata
\tablecomments{
Here, the PB20 data selection is identical to what is used in the paper, but the categorization between observation efficiency and data selection is modified, thus the values are different from those of Table 2 of \PBXX{}.}
\end{deluxetable}
\begin{deluxetable}{lrrr}
\tabletypesize{\footnotesize}
\tablecaption{Data Selection Efficiency\label{tab:dataselection}}
\tablehead{ & & & \colhead{Fractional}\\[-10pt]
\colhead{Stage of Data Selection} & \colhead{New} & \colhead{PB$20'$} & \colhead{Change}}
\startdata
Terminated/stuck observation & 98.8\% & 98.8\% & 0.0\% \\
Detector stage temperature & 98.8\% & 98.8\% & 0.0\% \\
Weather condition & 90.9\% & 90.8\% & 0.0\% \\
\begin{minipage}{4cm}Instrumental problem\\
\hspace*{1em}and volcano eruption\end{minipage} & 92.8\% & 92.6\% & $+0.2$\% \\
Data acquisition problem & 98.7\% & 98.9\% & $-0.2$\% \\
Calibration problem & 98.3\% & 98.3\% & 0.0\% \\
Off detector & 73.2\% & 73.7\% & $-0.6$\% \\
Individual detector glitch & 93.5\% & 93.5\% & 0.0\% \\
Common-mode glitch & 92.3\% & 60.6\% & $+52.4$\% \\
Remaining subscan fraction & 98.6\% & 80.5\% & $+22.6$\% \\
Individual detector PSD & 84.2\% & 80.8\% & $+4.2$\% \\
Common-mode PSD & 93.5\% & 93.0\% & $+0.5$\% \\
Map variance & 92.3\% & 95.4\% & $-3.3$\% \\
Low yield &100.0\% & 99.8\% & $+0.2$\% \\\hline Cumulative data selection & 36.2\% & 19.2\% & $+88.7$\%
\enddata
\tablecomments{
Since there are some updates in data selection, we cannot directly compare data selections in this work and \PBXX{}. Here, the PB20 data selection is reproduced applying the new data selection to PB20 intermediate data just for comparison, thus the cumulative efficiency is different from \cref{tab:datavolume}. }
\end{deluxetable}

The total observation time has slightly increased from \PBXX{} by 3.1\%. 
We use observations from 2014 July 25 until 2016 December 30. We also recovered some missing observations by reconstructing the database of all observations.

The detectors and their calibrations (pointing offset, relative gain, and relative polarization angle) used in this analysis are identical to those in \PBXX{}. Thus, we have no increase in detector yields.

The main data increase comes from improvements in data selection efficiency.
To explain how the encoder correction affects the data selection, we briefly explain our data selection procedure.

The data selection is done by flagging bad data whose data quality exceeds some threshold.
We use various types of data qualities evaluated using detector data, calibration data, housekeeping data, and external data.
Here, some data quality evaluations require data selection based on other low-level data qualities.
For example, 
bad sections of detector timestreams with glitches are masked in the evaluation of the power spectrum density (PSD) to prevent spurious contamination.

The data quality items and the requirements in the new analysis are almost the same as in
\PBXX{}. 
One property added is the fraction of remaining subscans.
Our one-hour observation consists of about 70 scans left and right at constant elevation. A subscan is each one-way scan.
The minimum unit of our data selection flagging is each subscan. Sometimes, most of the subscans are flagged, which causes a problem in the polynomial filter to detrend the baseline drift of the detector. In \PBXX{}, such data are removed in the timestream filtering step on the fly, and flagged as one of the reasons of failure in the evaluation of PSD.
In the new analysis, it is explicitly captured in the data selection.

\begin{figure}[!t]
\epsscale{1.15}
\plotone{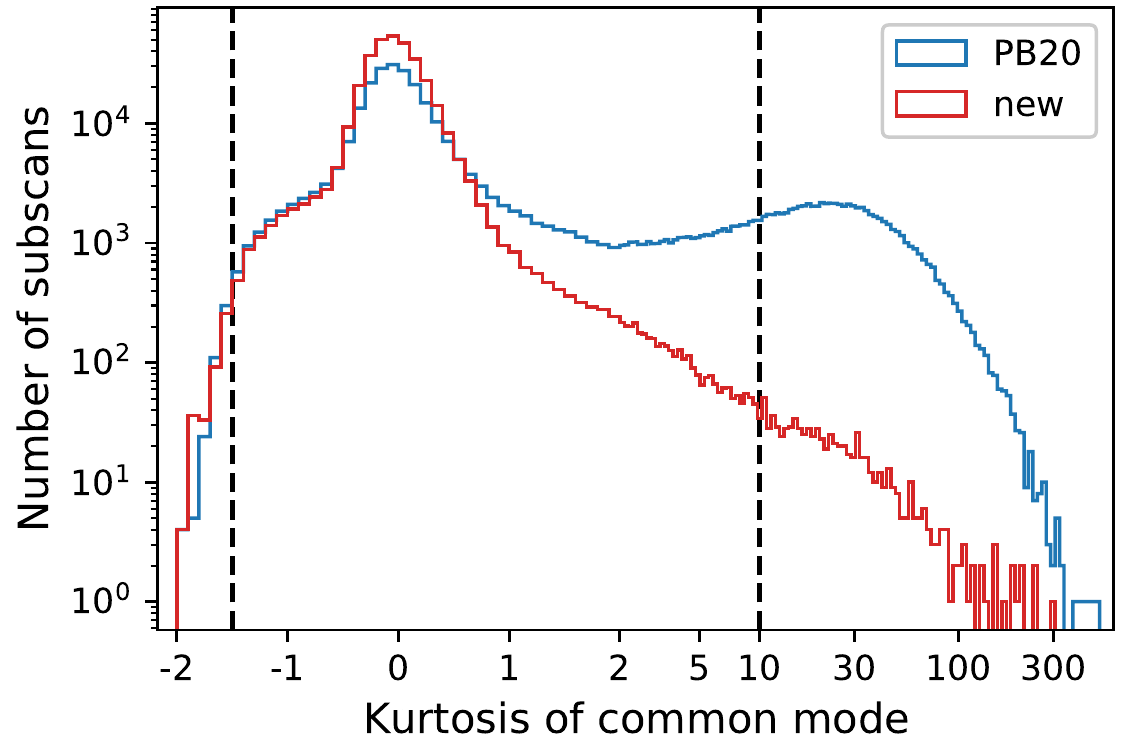}
\caption{Distribution of kurtosis computed for the common mode of the detector timestream. The blue line shows \PBXX{} data, and the red line shows new one with the encoder correction. The dashed lines are upper and lower thresholds for the data selection. When the timestream follows the normal distribution, the kurtosis becomes 0. Glitches make the kurtosis $>0$ as in the \PBXX{} data. Kurtosis $<0$ is due to residual low-frequency noise.
\label{fig:kurtosis}}
\end{figure}

Even with the same criteria for the data selection, the efficiencies improve for good-quality data.
The glitches due to the encoder error are detected in the common-mode glitch stage.
Here,
to test the Gaussianity of the averaged detector timestream $d_t$, we compute its kurtosis $K = {\left\langle(d_t-\langle d_t \rangle)^4\right\rangle}
/{\left\langle(d_t-\langle d_t \rangle)^2\right\rangle^2} - 3 $ and require $-1.5 < K < 10$ as shown in \Cref{fig:kurtosis}.
Thanks to the encoder correction, the glitches in the data have disappeared, and thus the efficiency of the step has significantly improved from 60.6\% to 92.3\%, which means an increase in the remaining data by 52.4\%.

In addition, 
the selection efficiency based on the remaining subscan fraction has also improved by 22.6\%. Sometimes, the encoder error occurs so often that most of subscans in a one-hour observation are flagged, then we have to drop the observation entirely. As the former selection efficiency based on glitches increases, the efficiency of this following data selection also improves.

\section{Analysis pipeline} \label{sec:pipeline}
\begin{figure*}[!t]
\plotone{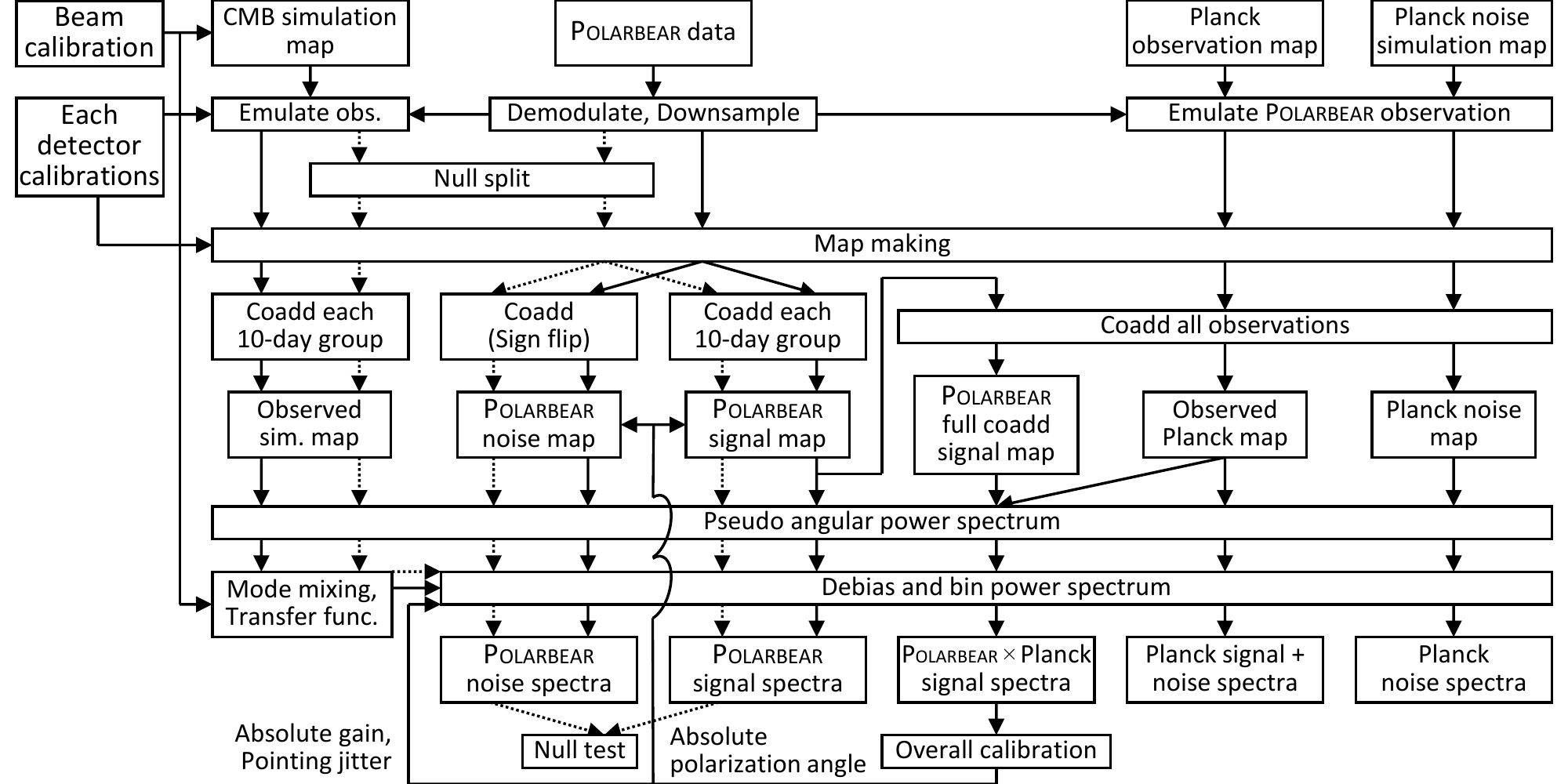}
\caption{Schematic view of the steps in our power spectrum estimation pipeline.
Each box represents an analysis procedure, input data, or its product.
The arrows show the flow of the pipeline, where the dotted arrows show the flow for the null test especially.
All the details of the pipeline are described in \PBXX{}.
The main change in this work is the update of the input \textsc{Polarbear} data to the new data set, as described in \cref{sec:glitch_dataselection}. 
The other procedures of the pipeline are almost the same as in \PBXX{}, except for the minor change in the overall calibration, as described in the text.
\label{fig:pipeline}}
\end{figure*}

Here, we analyze the new data set and measure the angular-power spectrum of the CMB $B$-mode at the multipole range of $50 < \ell < 600$.
\Cref{fig:pipeline} shows the overview of the analysis pipeline.
All the details are described in \PBXX{}. Here, we only report updates on the data validation, overall calibrations, and measured power spectra.

\subsection{
Calibrations Based on the Power Spectrum} \label{sec:calibration}
As described in \PBXX{}, we perform calibrations of the instruments in two steps. In the first step, we use calibration measurements, e.g., the thermal source calibrator and observations of planets and other bright sources, including Tau A. We calibrate the pointing model of the telescope, the pointing offsets of detectors, the effective beam function, relative gain variations, detector time constants, relative polarization angles, and polarization efficiencies. All these calibrations are the same as in
\PBXX{}. In the second step, we determine overall calibration parameters using the measured CMB angular-power spectra. We calibrate the absolute gain and the beam uncertainty using the $E$-mode autospectrum, and the absolute polarization angle using the $EB$ cross-spectrum.

The calibration error on power spectra due to the overall gain $g_0$ and beam uncertainty $\sigma^2$ is modeled as
\begin{equation}
{g}_\ell = g_0 \exp\left[-\frac{\ell(\ell + 1)} {2} \sigma^2\right]\;,
\end{equation}
where the exponential term represents Gaussian smearing due to the pointing jitter. 
Here, we treat $\sigma^2$ as a single parameter and allow it to be negative, which helps to remove a potential systematic error in our beam calibration.
We estimate $g_0$ and $\sigma^2$ using the Planck 2018 PR3 143\,GHz full mission maps.
We compute \textsc{Polarbear}-auto, Planck-auto, and \textsc{Polarbear}-cross-Planck $E$-mode binned power spectrum estimates,\footnote{We compute the angular-power spectra $C_\ell$ in the scaling of\\$D_\ell=\{\ell(\ell+1)/(2\pi)\} C_\ell$.}
$\hat{D}_{b,\mathrm{PB}}^{EE}$, $\hat{D}_{b,\mathrm{P143}}^{EE}$, and $\hat{D}_{b,\mathrm{PB}\times\mathrm{P143}}^{EE}$. The calibration errors in the \textsc{Polarbear} maps modify the spectra as
\begin{equation}
\left\{
\begin{array}{lcrl}
\hat{D}_{b,\mathrm{P143}}^{EE} & = & \tilde{D}_{b}^{EE} & + \Delta\hat{D}_{b,\mathrm{P143}}^{EE}\;, \\
\hat{D}_{b,\mathrm{PB}\times\mathrm{P143}}^{EE} & = & \hat{g}_{b} \tilde{D}_{b}^{EE} & + \Delta\hat{D}_{b,\mathrm{PB}\times\mathrm{P143}}^{EE}\;, \\
\hat{D}_{b,\mathrm{PB}}^{EE} & = & \hat{g}_{b}^2 \tilde{D}_{b}^{EE} & + \Delta\hat{D}_{b,\mathrm{PB}}^{EE}\;, 
\end{array}\right.\label{eq:gaincalCb1}
\end{equation}where $\tilde{D}_{b}^{EE}$ is the true $E$-mode spectra in the \textsc{Polarbear} patch, 
$\Delta\hat{D}_{b}^{EE}$ is the statistical uncertainty for each spectrum, and $\hat{g}_{b} = \sum_\ell w_{b\ell} g_\ell$ is the binned calibration factor weighted by the bandpass window function $w_{b\ell}$.
Here, we emulate \textsc{Polarbear} observations on the Planck maps and process the same map-making as \textsc{Polarbear} data. The resulting Planck map should contain the same signal $\tilde{D}_{b}^{EE}$ as \textsc{Polarbear}. Noise bias in the Planck autospectrum is estimated and subtracted using 96 realizations of Planck noise simulation maps. 
The
\textsc{Polarbear} autospectrum is calculated by cross-correlating 38 submaps grouped every 10 days, thus free from the noise bias. The uncertainties of the spectra are estimated using the noise simulations.

We fit \cref{eq:gaincalCb1} varying $g_0$, $\sigma^2$, and $\tilde{D}_b^{EE}$.
We find the overall gain calibration factor of $g_0 = 1.106 \pm 0.021$ and the beam uncertainty factor of $\sigma^2 = 0.99 \pm 3.00\,\mathrm{arcmin}^2$.
These are in good agreement with \PBXX{}, as expected.

We calibrate the absolute polarization angle using the observed $EB$ cross-spectrum. We assume that the original $EB$ correlation is null, but the observed one has finite values from the leakage of $E$-modes due to the angle error $\psi$ as~\citep{Minami2019PTEP}
\begin{equation}\label{eq:eb}
\hat{D}_{b,\mathrm{PB}}^{EB} = \frac{1}{2}\tan(4\psi)\, \hat{D}_{b,\mathrm{PB}}^{EE} + \Delta\hat{D}_{b,\mathrm{PB}}^{EB}\;.
\end{equation}
Here, we use the observed $E$-mode spectra instead of the theoretical one used in \PBXX{}.
It makes the result independent of the absolute gain calibration and the fiducial cosmological parameters.
The uncertainty of the spectrum $\Delta\hat{D}_{b,\mathrm{PB}}^{EB}$ is estimated using the quasi-analytic method (\PBXX{}) as
\begin{equation}
\Delta\hat{D}_{b,\mathrm{PB}}^{EB} = \sqrt{
\frac{(\hat{D}_{b,\mathrm{PB}}^{EE} + N_{b,\mathrm{PB}}^{EE})N_{b,\mathrm{PB}}^{EE}}{\nu_{b,\mathrm{PB}}^{EE}}
}\;,
\end{equation}
where $N_{b,\mathrm{PB}}^{EE}$ is the noise bias of \textsc{Polarbear} autospectrum estimated using the noise simulations, and $\nu_{b,\mathrm{PB}}^{EE}$ is the degrees of freedom, estimated as
\begin{equation}
\nu_{b,\mathrm{PB}}^{EE} = 2 \left(N_{b,\mathrm{PB}}^{EE}/\Delta\hat{D}_{b,\mathrm{PB}}^{EE}\right)^2\;.
\end{equation}
Here we assume that the noise bias and degrees of freedom for $B$-modes are similar to $E$-modes.

By fitting \cref{eq:eb}, we find our angle error $\psi=-0\fdg67\pm0\fdg15$, 
which is also consistent with the previous analysis. 

\begin{deluxetable}{lrrr}
\tablecaption{Null Test Total $\chi^2$ PTE Values \label{tab:nullchi2}}
\tabletypesize{\scriptsize}
\tablehead{
\colhead{} & \colhead{$EE$} & \colhead{$EB$} & \colhead{$BB$}
} 
\startdata
Null test summed over $\ell$ bins \\
First half versus second half 	& 0.6\% & 67.8\% & 84.0\% \\
Rising versus middle and setting 	& 1.2\% & 7.4\% & 2.2\% \\
Middle versus rising and setting 	& 10.0\% & 54.2\% & 78.4\% \\
Setting versus rising and middle 	& 78.6\% & 50.8\% & 76.6\% \\
Left-going versus right-going subscans 	& 0.4\% & 54.2\% & 4.8\% \\
High-gain versus low-gain CESs 	& 15.6\% & 82.4\% & 81.8\% \\
High PWV versus low PWV 	& 30.6\% & 47.6\% & 74.8\% \\
Common-mode $Q$ knee frequency 	& 64.2\% & 88.4\% & 36.0\% \\
Common-mode $U$ knee frequency 	& 67.0\% & 65.0\% & 17.6\% \\
Mean temperature leakage by bolometer 	& 62.8\% & 24.6\% & 50.4\% \\
$2f$ amplitude by bolometer 	& 60.6\% & 65.0\% & 16.6\% \\
$4f$ amplitude by bolometer 	& 59.6\% & 90.0\% & 30.4\% \\
$Q$ versus $U$ pixels 	& 84.0\% & 8.2\% & 17.0\% \\
Sun above or below the horizon 	& 9.8\% & 56.6\% & 57.4\% \\
Moon above or below the horizon 	& 88.8\% & 14.0\% & 50.2\% \\
Top half versus bottom half 	& 84.8\% & 97.8\% & 41.0\% \\
Left half versus right half 	& 85.0\% & 2.0\% & 94.8\% \\
Top versus bottom bolometers 	& 50.2\% & 71.6\% & 47.6\% \\
\hline
$\ell$ bin summed over null tests \\
$50 \le \ell \le 100$ & 7.4\% & 87.0\% & 37.4\% \\
$100 < \ell \le 150$ & 53.4\% & 74.8\% & 49.4\% \\
$150 < \ell \le 200$ & 97.6\% & 59.2\% & 25.8\% \\
$200 < \ell \le 250$ & 95.4\% & 17.4\% & 5.4\% \\
$250 < \ell \le 300$ & 85.4\% & 73.6\% & 33.6\% \\
$300 < \ell \le 350$ & 12.2\% & 70.0\% & 97.6\% \\
$350 < \ell \le 400$ & 0.0\% & 83.8\% & 85.0\% \\
$400 < \ell \le 450$ & 25.8\% & 10.6\% & 11.2\% \\
$450 < \ell \le 500$ & 58.6\% & 2.2\% & 15.4\% \\
$500 < \ell \le 550$ & 70.8\% & 99.4\% & 53.0\% \\
$550 < \ell \le 600$ & 0.2\% & 17.4\% & 89.6\% \\
\enddata
\vspace*{-20pt}
\end{deluxetable}\begin{deluxetable}{lrrr}
\tablecaption{Null Test PTE Values \label{tab:nullstats}}
\tablehead{
\colhead{Null Statistic} & \colhead{$EE$} & \colhead{$EB$} & \colhead{$BB$}
} 
\startdata
Average $\chi$ overall & 93.4\% & 36.4\% & 92.4\% \\
Most extreme $\chi^2$ by bin & 3.6\% & 24.0\% & 45.8\% \\
Most extreme $\chi^2$ by test & 8.2\% & 38.0\% & 32.0\% \\
Most extreme $\chi^2$ overall & 13.8\% & 42.2\% & 68.0\% \\
Total $\chi^2$ overall & 8.6\% & 56.6\% & 33.8\% \\
\hline
Lowest statistic & 14.0\% & 61.2\% & 75.8\% \\
\hline
KS test on all bins & 47.5\% & 45.4\% & 75.3\% \\
KS test on all spectra & 57.3\% & 42.8\% & 98.6\% \\
KS test overall	 & 19.8\% & 98.4\% & 72.4\% \\
\enddata
\vspace*{-20pt}
\end{deluxetable}
\vspace*{-2\intextsep}
\subsection{Data Validations} \label{sec:validation}
As described in \cref{sec:glitch_dataselection}, the main difference of the new data set from \PBXX{} is the correction of the glitch due to the HWP angle error. It should not inject any systematic uncertainties. Therefore, we use the same estimates of the systematic uncertainties done in \PBXX{}. However, thanks to the increase of statistics, unknown systematics may become noticeable. Thus, we perform the null tests in \PBXX{} with the new data set.

The null tests are performed for the same 18 splits as \PBXX{}.
The null spectra are compared to 500 noise-only simulations generated using the ``signflip'' method.
We estimate the uncertainty of the null spectra from the noise simulations and evaluate a $\chi$ value for each spectrum, each null split, and each $\ell$ bin. We compute the sum of $\chi^2$ over $\ell$ bins or over null splits. Then we evaluate the probability to exceed (PTE) values by counting the fraction of the noise simulations whose total $\chi^2$ exceeds the value of real data. \Cref{tab:nullchi2} shows the results. 
Next, we compute five representative statistics: (1) the average of $\chi$ among all $\ell$ bins and all null splits, (2) the most extreme total $\chi^2$ by $\ell$ bin summed over null splits, (3) the most extreme total $\chi^2$ by test summed over $\ell$ bins, (4) the most extreme $\chi^2$ among all $\ell$ bins and all null splits, and (5) the total $\chi^2$ summed over $\ell$ bins and null splits. The PTEs are computed by comparing the statistics from real data with those evaluated from every realization of the noise simulations.
Finally, we choose the lowest PTE of the five statistics and evaluate its PTE again by comparing with the lowest PTEs from the noise simulations. \Cref{tab:nullstats} shows the results. In the $EE$ spectrum, for example, the lowest PTE is 3.6\%, but in 14.0\% of the noise simulations, the lowest PTE becomes lower than 3.6\%. Thus, we obtain the final PTE of 14.0\%.
In addition, we test the consistency of the distribution of PTE estimates with a uniform distribution using a Kolmogorov--Smirnov (KS) test. We test distributions of PTEs in \cref{tab:nullchi2}, as well as PTEs for individual $\chi$ values. The results are shown in \cref{tab:nullstats}.
We require that the PTE values of the lowest statistic and KS tests must be greater than 5\%.
All spectra ($EE$, $EB$, and $BB$) pass these criteria, as shown in \cref{tab:nullstats}.

\begin{figure}
\epsscale{1.15}
\plotone{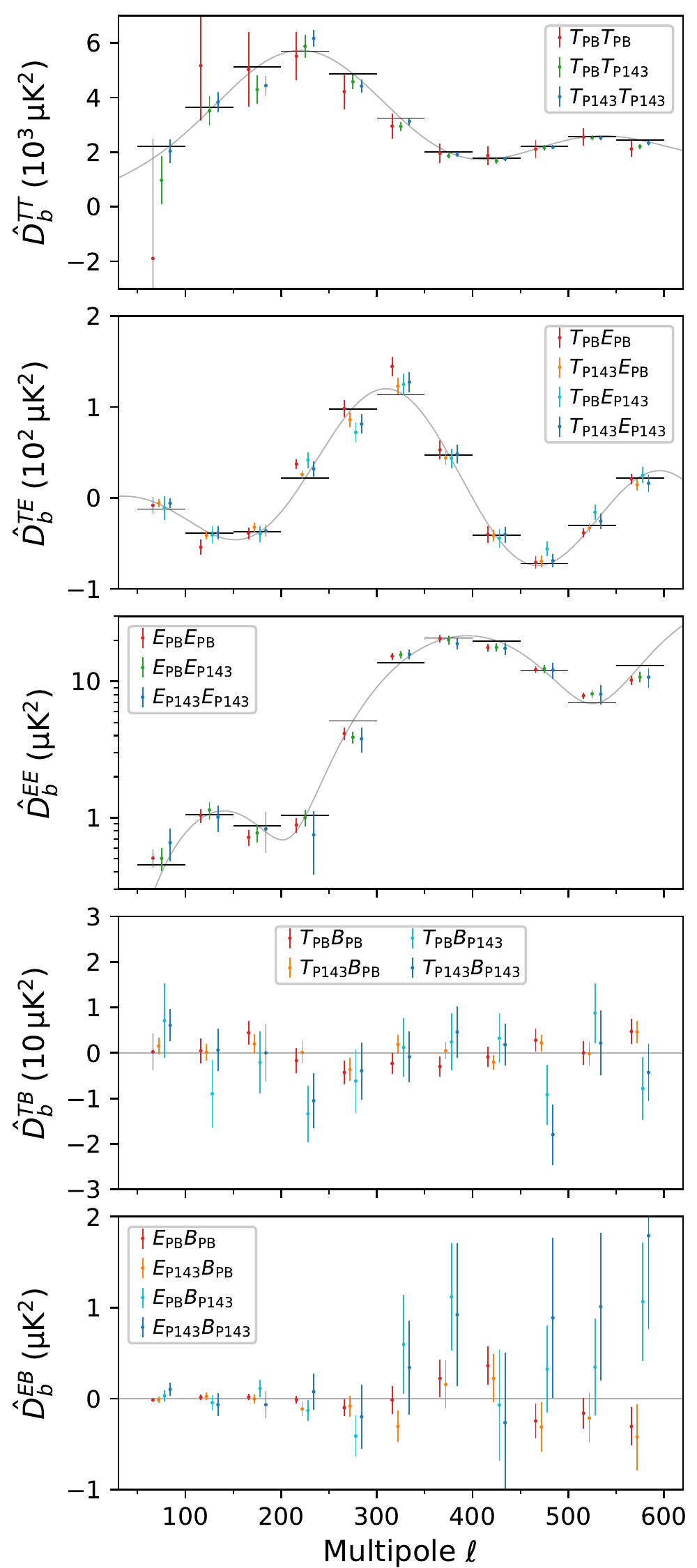}
\caption{
Binned estimates for the angular-power spectrum for $TT$, $TE$, $EE$, $TB$, and $EB$ spectra from the top to the bottom panel. The red and blue points show the spectra from \textsc{Polarbear} maps and Planck $143\,\mathrm{GHz}$ maps. The orange, green, and cyan points show the spectra from different combinations of cross correlation between \textsc{Polarbear} and Planck.
The error bars include the sample variance.
Gray lines show the theoretical estimates with our fiducial cosmological parameters, and the black points are their binned averages. \label{fig:otherspectra}}
\end{figure}

Finally, we check the power spectra except for $B$-mode autocorrelation. As absolute calibrations in \cref{sec:calibration}, we use Planck $143\,\mathrm{GHz}$ maps as reference. \Cref{fig:otherspectra} shows the power spectra from \textsc{Polarbear} and Planck maps.
We take the inverse-variance weighted average\footnote{Here, we use the variance without the sample variance. } of the three or four auto- and cross spectra and compute the total $\chi^2$ for each spectrum compared to the averaged one. \Cref{tab:PBvsPlanck} shows the PTE values of the total $\chi^2$ as well as that of the overall total $\chi^2$.
We find that all these spectra are consistent between \textsc{Polarbear} and Planck.
Here, we apply the overall calibrations, which use $EE$ and $EB$ spectra as \cref{sec:calibration}. The consistencies of $TT$, $TE$, and $TB$ spectra support the robustness of our calibrations and analysis methods. 
\begin{deluxetable}{crrrrr}
\tablecaption{PTE Values of the Planck Consistency Check\label{tab:PBvsPlanck}}
\tablehead{
\colhead{Combination}
& \colhead{$TT$}
& \colhead{$TE$}
& \colhead{$EE$}
& \colhead{$TB$}
& \colhead{$EB$}
}
\startdata
\textsc{Polarbear} auto & 98.7\% & 2.8\% & 94.2\% & 84.8\% & 92.4\% \\
\textsc{Polarbear}$\times$Planck & & 14.4\% & & 24.6\% & 20.5\% \\[-5pt]
& 35.5\% & & 97.8\% & & \\[-5pt] Planck$\times$\textsc{Polarbear} & & 14.8\% & & 73.1\% & 88.6\% \\
Planck $143\,\mathrm{GHz}$ auto & 30.3\% & 98.2\% & 98.9\% & 19.0\% & 35.7\% \\\hline Overall & 65.0\% & 13.2\% & 99.7\% & 50.4\% & 69.8\%\enddata
\end{deluxetable}
\subsection{Results of $B$-mode Power Spectrum Estimates}
\begin{figure}[!t]
\epsscale{1.15}
\plotone{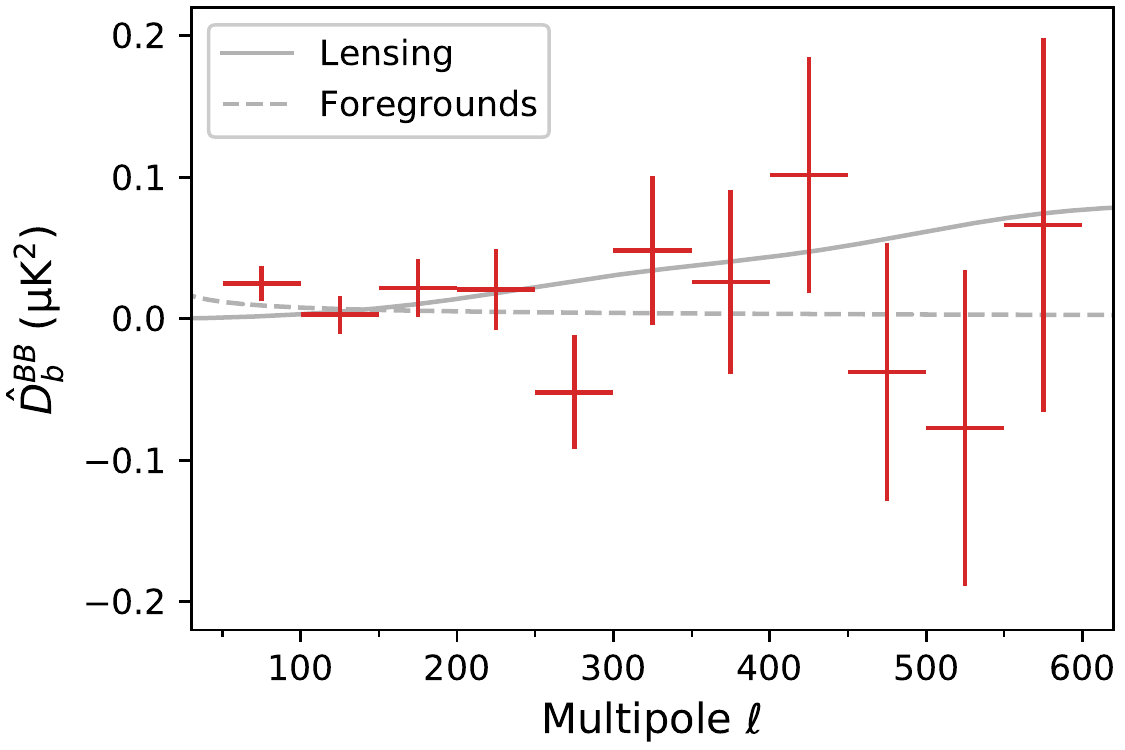}
\caption{Binned estimate of the angular-power spectrum for the $B$-mode from \textsc{Polarbear} maps. The error bars include only the statistical uncertainties. The solid gray line shows the theoretical estimate of the lensing $B$-mode. The dashed line shows the power-law model of contaminations due to Galactic dust foregrounds. \label{fig:Bspec}}
\end{figure}

The results of $B$-mode power spectrum measurements with the new data set are summarized in \cref{tab:Bspec} and shown in \cref{fig:Bspec}. Again, we estimate the statistical uncertainties using the quasi-analytic method with 500 noise-only simulations based on the ``signflip'' method. The estimates of systematic uncertainties are same with \PBXX{} because we use the same period of observations and the same instrumental calibrations. The overall calibration uncertainties are multiplicative error due to the overall gain and beam calibrations in \cref{sec:calibration}.
\begin{deluxetable}{rrrrr}
\tablecaption{$B$-mode Band Powers and Uncertainties \label{tab:Bspec}}
\tablehead{
\colhead{Band Definition}
& \colhead{$\hat{D}_{b}^{BB}$}
& \colhead{$\Delta\hat{D}_{b,\mathrm{stat}}^{BB}$}
& \colhead{$\Delta\hat{D}_{b,\mathrm{syst}}^{BB}$}
& \colhead{$\Delta\hat{D}_{b,\mathrm{cal}}^{BB}$}
\\[-8pt]
\colhead{}
& \colhead{($\mu\mathrm{K}^2$)}
& \colhead{($\mu\mathrm{K}^2$)}
& \colhead{($\mu\mathrm{K}^2$)}
& \colhead{($\mu\mathrm{K}^2$)}
}
\startdata
$ 50 < \ell \leq 100$ & 0.0249 & 0.0126 & 0.0040 & 0.0010 \\
$100 < \ell \leq 150$ & 0.0029 & 0.0135 & 0.0010 & 0.0001 \\
$150 < \ell \leq 200$ & 0.0218 & 0.0207 & 0.0012 & 0.0009 \\
$200 < \ell \leq 250$ & 0.0207 & 0.0287 & 0.0007 & 0.0008 \\
$250 < \ell \leq 300$ & $-$0.0521 & 0.0403 & 0.0015 & 0.0022 \\
$300 < \ell \leq 350$ & 0.0481 & 0.0528 & 0.0022 & 0.0022 \\
$350 < \ell \leq 400$ & 0.0259 & 0.0650 & 0.0022 & 0.0013 \\
$400 < \ell \leq 450$ & 0.1016 & 0.0835 & 0.0021 & 0.0059 \\
$450 < \ell \leq 500$ & $-$0.0376 & 0.0912 & 0.0029 & 0.0025 \\
$500 < \ell \leq 550$ & $-$0.0772 & 0.1114 & 0.0019 & 0.0060 \\
$550 < \ell \leq 600$ & 0.0664 & 0.1323 & 0.0049 & 0.0060 \\
\enddata
\tablecomments{$\hat{D}_{b}^{BB}$ is the binned angular-power spectrum estimates from the \textsc{Polarbear} maps. $\Delta\hat{D}_{b,\mathrm{stat}}^{BB}$, $\Delta\hat{D}_{b,\mathrm{syst}}^{BB}$, and $\Delta\hat{D}_{b,\mathrm{cal}}^{BB}$ are uncertainties on the band powers due to noise statistics, instrumental systematics, and overall calibration uncertainties, respectively. We assume the same estimates of $\Delta\hat{D}_{b,\mathrm{syst}}^{BB}$ as \PBXX{}.}
\vspace*{-20pt}
\end{deluxetable}

We compare our $B$-mode measurements with a model of $B$-mode signal based on the Planck 2018 $\Lambda$CDM lensing $B$-mode spectrum and a foreground component modeled by \cite{BK15}. Direct comparison gives the reduced $\chi^2$ of 9.34/11, indicating good agreement.
Fitting an overall amplitude rescaling this template, we find $A_{BB} = 0.59^{+0.46}_{-0.31}$, and the null hypothesis is disfavored at $1.4\,\sigma$. 
Note that this significance is lower than the $2.2\,\sigma$ in \PBXX{}, but the uncertainty on $A_{BB}$ has improved from $0.8$ thanks to the increase of data volume. See \cref{sec:comparison} for more detailed discussions about the consistency with \PBXX{}.

In \cref{sec:pe} we perform more detailed parameter estimations combining Planck 2018 PR3 observations at 143, 273, and $353\,\mathrm{GHz}$.
As we do for the $143\,\mathrm{GHz}$ map used for the absolute gain calibration in \cref{sec:calibration}, we emulate \textsc{Polarbear} observations on the Planck maps, and process map-making as for
\textsc{Polarbear} data.
We stack maps from all emulated observations by frequency, and compute auto- and cross spectra from these Planck three-frequency maps as well as from the \textsc{Polarbear} map.
Here, auto-spectra from the same Planck observation contain the noise bias, which is estimated using the 96 Planck noise simulation maps for each frequency.

\section{Parameter Estimation} \label{sec:pe}
In this section we briefly explain our likelihood, the constraints on the tensor-to-scalar ratio $r$, and the foreground contamination in our $BB$ spectrum. We follow the same procedure as \PBXX{}, 
which we summarize briefly here.
Our estimation uses $BB$ signal and noise spectra of four auto- and six cross spectra from \textsc{Polarbear} $150\,\mathrm{GHz}$, Planck $143\,\mathrm{GHz}$, Planck $273\,\mathrm{GHz}$, and Planck $353\,\mathrm{GHz}$.
We arrange these 10 measured signal spectra in such a way that each bandpower is a matrix $\hat{D}_{b,ij}$, where $i$ and $j$ are one of the four observations. The diagonal block of this matrix contains four autospectrum values, and the off-diagonal block contains six cross-spectrum values.
The noise spectra are also arranged as $N_{b,ij}$.
Since noise between frequency bands and experiments is expected to be uncorrelated, $N_{b,ij}$ is diagonal.

The signal spectra $D_{b,ij}$ can be modeled as a sum of CMB and foregrounds as
\begin{equation}
D_{b,ij} = D_{b}^{\mathrm{CMB}} + D_{b,ij}^{\mathrm{fg}}\;,
\label{eq:total_signal_fg}
\end{equation}
where $D_{b,ij}^{\mathrm{fg}}$ is the total foreground component, which depends on frequencies.
The CMB component $D_{b}^{\mathrm{CMB}}$ is the same among the frequencies and has contributions from primordial $B$-modes and from the weak lensing of the CMB. Thus it can be modeled as
\begin{equation}
D_{b}^{\mathrm{CMB}} = r D_{b}^{\mathrm{tens}} + A_{\mathrm{lens}} D_{b}^{\mathrm{lens}}\;,
\end{equation}
where $r$ is the tensor-to-scalar ratio, and $A_{\mathrm{lens}}$ is the normalized amplitude of the $\Lambda$CDM lensing signal. $D_{b}^{\mathrm{tens}}$ and $D_{b}^{\mathrm{lens}}$ are the binned tensor and lensing signals, respectively.
The tensor spectrum at $r=1$ is computed using \texttt{CAMB}~\citep{Lewis:1999bs} with a tensor power amplitude of $2.46 \times 10^{-9}$ and a spectral index of zero.
As shown in \PBXX{}, the upper limit of contamination from the polarized sources and synchrotron emission is subdominant to the dust component at $150\,\mathrm{GHz}$. So we assume that our measured spectra $\hat{D}_{b,ij}$ are contaminated by foregrounds mostly from Galactic dust, $D_{b,ij}^{\mathrm{fg}} \approx D_{b,ij}^{\mathrm{dust}}$,
where $D_{b,ij}^{\mathrm{dust}}$ is the dust component, which we consider a power law in $\ell$ and a modified blackbody in frequencies $i$, $j$ as defined in \citet{p2016} and \citet{P2018e},
\begin{equation}
D_{b,ij}^{\rm dust} = \sum_{\ell} w_{b\ell}\, A_{\mathrm{dust}}\,f_i\,f_j\,\left( \frac{\ell}{\ell_0} \right)^{\alpha_\mathrm{dust}}\;.
\end{equation}
Here $w_{b\ell}$ is the bandpass window function. 
$A_{\mathrm{dust}}$ is the amplitude of the dust component, and $\alpha_{\mathrm{dust}}$ is the power-law index in $\ell$. We consider a pivot value of $\ell_0 = 80$. The $f_{i}$ is the dust emission at each frequency bandpass in CMB temperature units defined as $f(\beta_{\mathrm{dust}},T_{\mathrm{dust}})$, where $\beta_{\mathrm{dust}}$ is the spectral index, and $T_{\mathrm{dust}}$ is the temperature of the modified blackbody. The $f_i$ is normalized such that $A_{\mathrm{dust}}$ corresponds to the dust emission at $353\,\mathrm{GHz}$.

\subsection{Likelihood}
Under the assumption according to \citet{hl08} that the measured $\hat{D}_{b,ij}^{\mathrm{tot}}=\hat{D}_{b,ij} + N_{b,ij}$ follows a Wishart distribution~\citep{Wishart1928} with an effective number of degrees of freedom $\nu_b$, we define our likelihood $\mathcal{L}$ of the true spectrum $D_{b,ij}^{\mathrm{tot}}=D_{b,ij} + N_{b,ij}$ given the measured $\hat{D}_{b,ij}^{\mathrm{tot}}$ as
\begin{equation}
\begin{array}{l}
-2 \; \ln \; \mathcal{L} = \sum_b \nu_b \Big\{\\
\quad\quad\text{Tr} \left[\hat{D}_{b}^{\mathrm{tot}} \left({D}_{b}^{\mathrm{tot}}\right)^{-1}\right] - \ln \left|\hat{D}_{b}^{\mathrm{tot}} \left({D}_{b}^{\mathrm{tot}}\right)^{-1}\right| - n_{\mathrm{freq}} \Big\}\;.
\end{array}
\end{equation}
The effective number of degrees of freedom $\nu_b$ is estimated from the standard deviation of the spectrum of the noise realizations. The standard deviation in auto- and cross-spectrum is determined following \PBXX{} as
\begin{equation}
\nu_{b,i} = 2 \left( \frac{N_{b,ii}}{\sigma(N_{b,ii})}\right)^2\;.
\end{equation}
For our estimation, we use the geometrical mean of $\nu_b$ of \textsc{Polarbear} and Planck.

\begin{figure*}[t!]
\epsscale{0.7}
\plotone{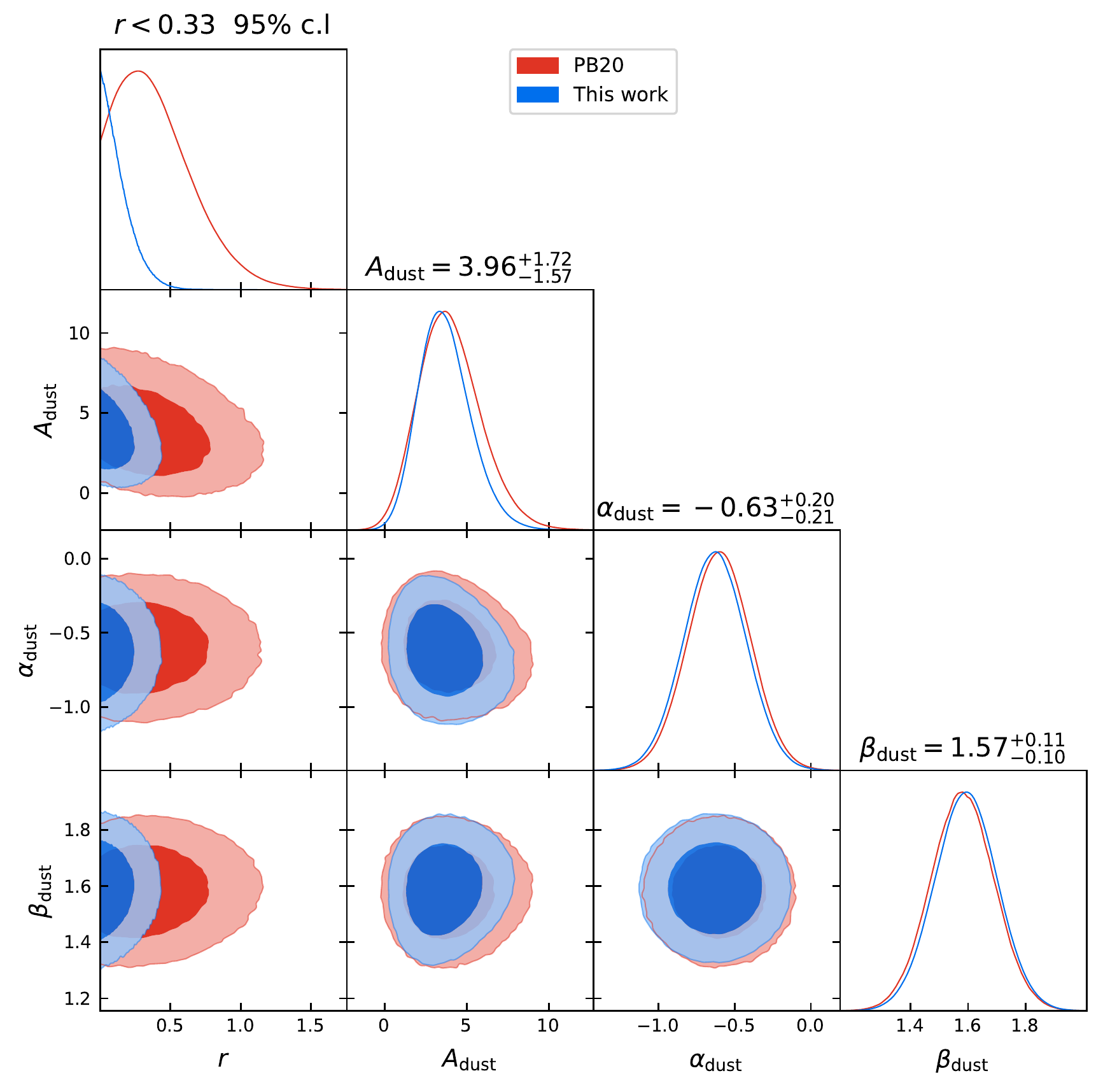}
\caption{Marginal posteriors for the four parameters $r$, $A_\mathrm{dust}$, $\alpha_\mathrm{dust}$, and $\beta_\mathrm{dust}$. We compare our previous estimate (red) with the reanalysis estimate (blue). \label{fig:corner}}
\end{figure*}
We sample our likelihood using \texttt{EMCEE} \citep{emcee} for the parameter estimation. Our model contains four free parameters, $r$, $A_\mathrm{dust}$, $\alpha_\mathrm{dust}$, and $\beta_\mathrm{dust}$. We fixed the values of $A_\mathrm{lens} = 1$ and $T_\mathrm{dust}=19.6\,\mathrm{K}$ \citep{p15}. For $\alpha_\mathrm{dust}$ and $\beta_\mathrm{dust}$, we considered Gaussian priors $-0.58 \pm 0.21$ and $1.59 \pm 0.11$, respectively \citep{BK15,p2016}.

\subsection{Constraints on Parameters}

Marginalized 68\% and 95\% parameter constraint contours and the posteriors are shown in \cref{fig:corner}. Similar to \PBXX{}, the posteriors of $\alpha_\mathrm{dust}$ and $\beta_\mathrm{dust}$ are dominated by the priors because these parameters are much less sensitive to the \textsc{Polarbear} data. 
Our estimate excludes the zero dust foregrounds with 99\% confidence and shows evidence of dust $B$-modes with amplitude $A_\mathrm{dust} = 4.0_{-1.6}^{+1.7}$.
The parameter $A_\mathrm{dust}$ is mostly constrained by Planck data. The 10\% increase in the best-fit value compared to 
\PBXX{} is due to the degeneracy with the parameter $r$. We find a 68\% confidence level maximum likelihood value of $r = -0.04_{-0.15}^{+0.18}({\rm stat})\;\pm\;0.03({\rm sys})$. 
We report the improved 95\% confidence upper limit of $r<0.33$ after marginalizing over foreground parameters,
requiring $r$ and $A_\mathrm{dust}$ to be positive a posteriori.
The new addition of data tightens the constraint on $r$ by a factor of 2.7 compared to \PBXX{}. 
We validate the constraint in \cref{sec:comparison}.

\subsection{Goodness of Fit}
\begin{figure}[h!]
\epsscale{0.95}
\plotone{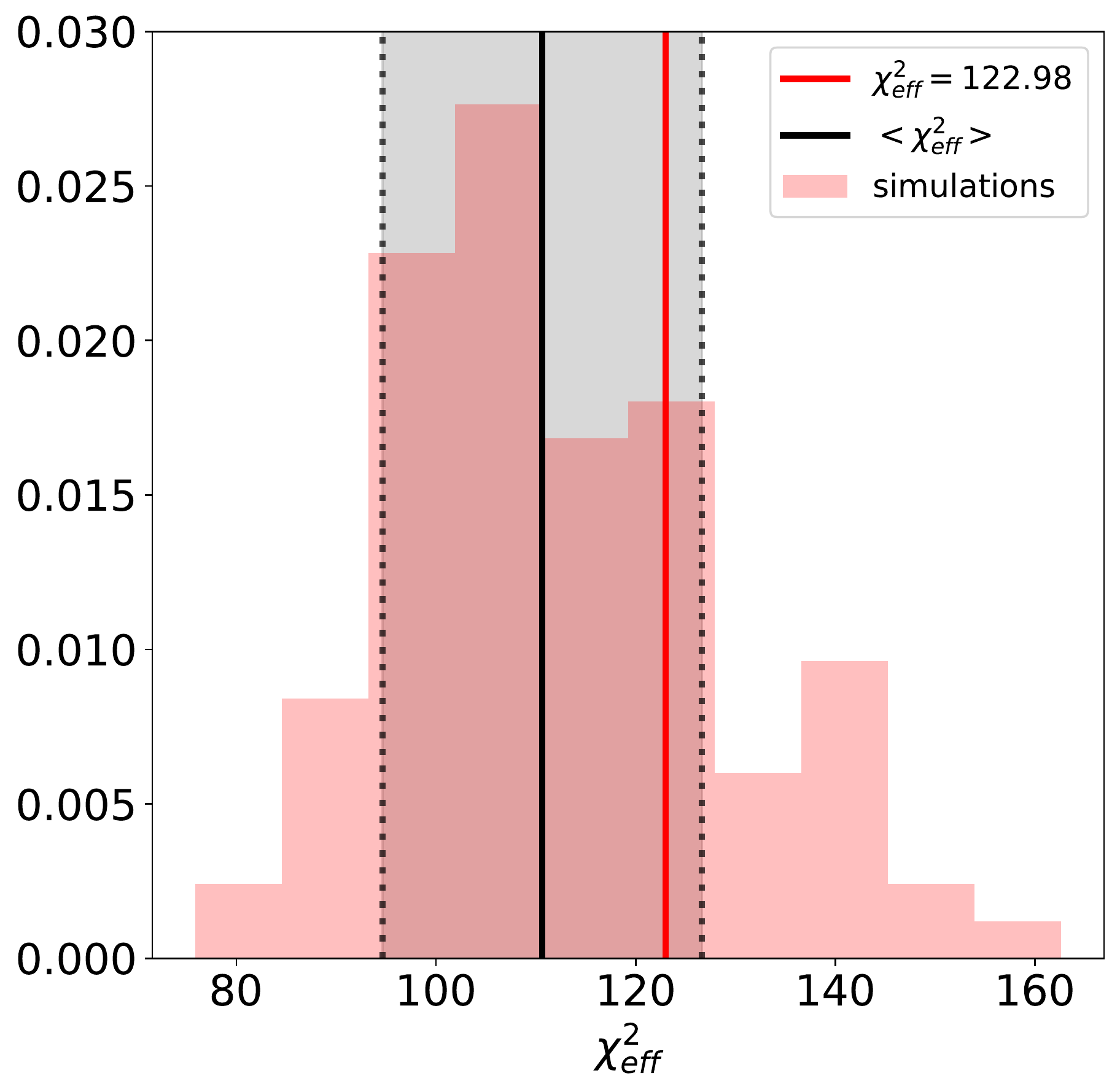}
\caption{Effective chi-squared of the best fit as the red vertical line. The black line is the expectation value of the Wishart distribution, \cref{eq:w_exp}, and the shaded area within the two dotted black verticals is the variance of the Wishart distribution, \cref{eq:w_var}. The histogram in red shows the effective chi-square distribution for a set of 96 simulations. \label{fig:chi}}
\end{figure}

\begin{figure}[h!]
\plotone{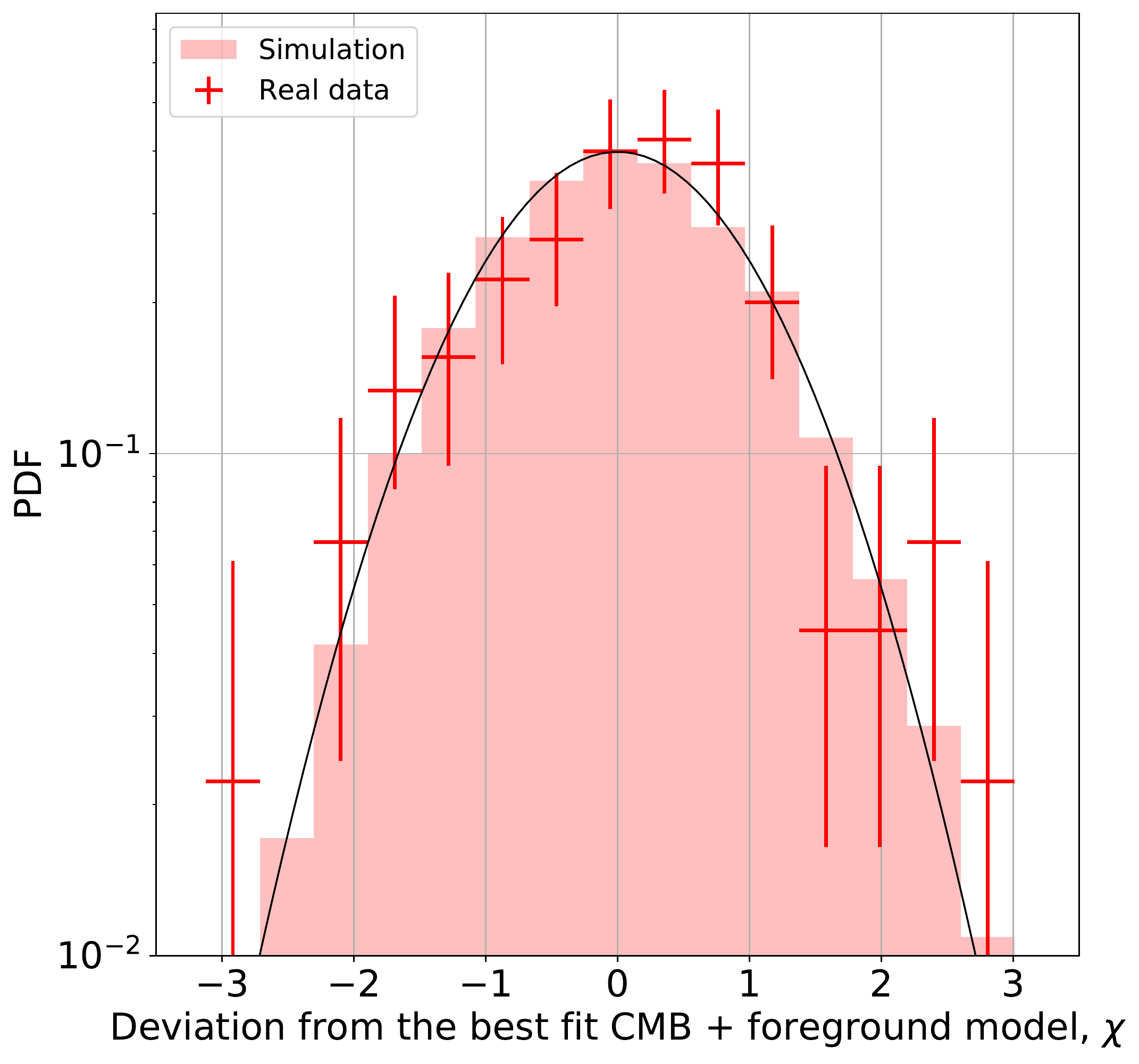}
\caption{Normalized difference between the measured cross spectra and the best-fit CMB+foreground model shown in units of standard deviation. The red error bars represent the real data, and the histogram is obtained from a set of 96 simulations. \label{fig:dev}}
\end{figure}

As a measure of the goodness of fit of $D_{b,ij}$ to $\hat{D}_{b,ij}$, we can define an effective chi-square following \citet{hl08},
\begin{equation}
\chi_\mathrm{eff}^2 = -2 \: \text{ln} \: \mathcal{L}\;.
\end{equation}
Here, 
$\chi_\mathrm{eff}^2$ is zero if $D_{b,ij} = \hat{D}_{b,ij}$. For $\nu_b \gg n_\mathrm{freq}$ and in the limit of a negligible number of fit parameters compared to the total number of bins across all spectra, the expectation value and variance of the effective chi-square under the Wishart distribution is given by
\begin{equation}
\left\langle \chi_\mathrm{eff}^2\right\rangle \approx n_\mathrm{bins}\frac{n_\mathrm{freq}(n_\mathrm{freq} + 1)}{2}\;,
\label{eq:w_exp}
\end{equation}
and
\begin{equation}
\text{Var}\left(\chi_{\mathrm{eff}}^2\right) \approx 2\left\langle \chi_{\mathrm{eff}}^2\right\rangle\;,
\label{eq:w_var}
\end{equation}
where $n_\mathrm{bins}$ is the number of multipole bins of the spectra. In \cref{fig:chi} we show our maximum likelihood $\chi_{\mathrm{eff}}^2 = 122.98$ with the red line. The solid vertical black line shows the expected value, and the gray shaded area shows the variance. The $\chi_{\mathrm{eff}}^2$ of our data is consistent with the simulated $\chi_{\mathrm{eff}}^2$, and it lies within the expected variance under the Wishart distribution. In \cref{fig:dev} we show the normalized difference between the measured cross spectra and the best-fit CMB+foregrounds model.


\section{Comparison with the Results of PB20} \label{sec:comparison}
Here, we compare the new results with those of \PBXX{} and discuss their consistency.
Since both data sets pass the null tests, the results are statistically valid.
However, the new data set significantly overlapsthe
\PBXX{} data set,
and the change should be the result of the additional data recovered in \cref{sec:dataselection}.

\begin{figure*}[t]
\epsscale{1.15}
\plotone{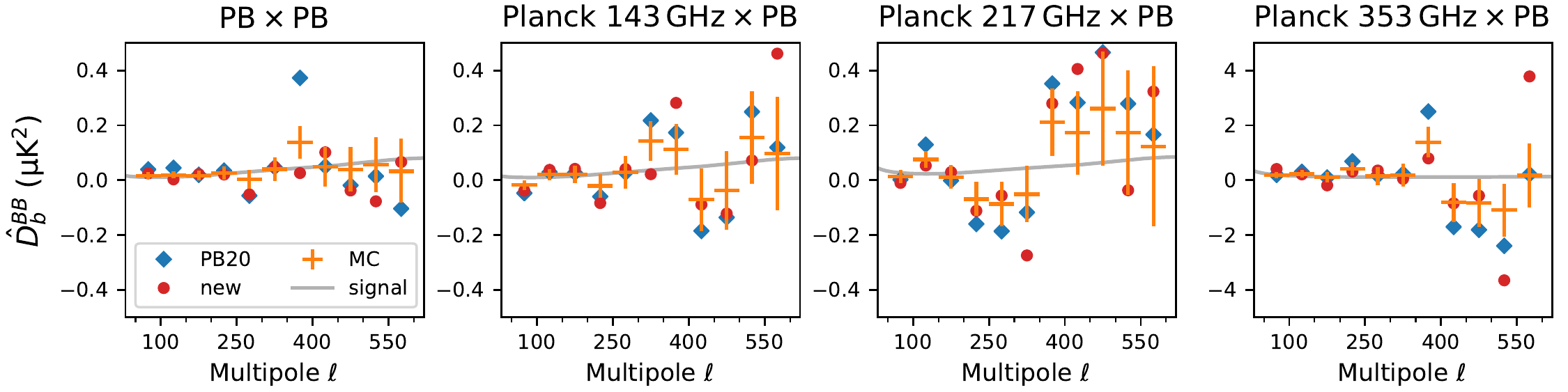}
\caption{MC simulations of the second measurement from \PBXX{} with additional data. The \textsc{Polarbear} autospectrum and cross spectra with Planck 143, 217, and $353\,\mathrm{GHz}$ are shown from left to right. The orange points show the median value and $1\sigma$ range of the MC simulations. 
The blue and red points are the central values of $\hat{D}_{b}^{BB}$ in \PBXX{} and in this work, respectively. The gray line shows the signal spectrum assumed for the MC simulations.\label{fig:MCps}}
\end{figure*}

In this section, we evaluate the probability of the consistency using a Monte Carlo (MC) simulation as follows. First, we evaluate the effective data increase between \PBXX{} and the new data set. Next, we simulate expected shifts of measurements from \PBXX{} due to the increase of statistics. Finally, we evaluate the probability by comparing the actual new measurement with the MC simulations.

\subsection{Effective Increase in the Data Volume}\label{sec:effadd}
First, we evaluate the effective data increase by comparing the noise power spectrum, $N_b$.
The estimate of fractional data increase of 81.6\% in \cref{sec:dataselection} assumes uniform weights among detectors and observations.
In practice, however, each observation has a different noise performance depending on the weather and other instrumental conditions.
In the map-making and coadding steps, we average data using the inverse-variance weighting.
If the recovered data have a better noise performance than the
\PBXX{} data set, the effective data increase can exceed the naive estimation.

In the analytic power spectrum uncertainty estimation known as the Knox formula~\citep{Knox1995},
the inverse of the noise power spectrum is proportional to the sensitivity-weighted data volume per solid angle as
\begin{equation}
\frac{1}{N_\ell} = \frac{n_\mathrm{det}t_\mathrm{obs}}{2N_\mathrm{NET}^2 4\pi f_\mathrm{sky}} \exp[-\ell(\ell+1)\sigma_\mathrm{beam}^2]\;,
\end{equation}
where $n_\mathrm{det}$ is the total number of detectors, $t_\mathrm{obs}$ is the total observation time, $N_\mathrm{NET}$ is the instantaneous sensitivity of a single detector, $4\pi f_\mathrm{sky}$ is the solid angle of the observing patch, and the exponential term is the Gaussian beam smearing effect.
We generalize this formula and apply it to our binned noise power spectrum, $N_b = \sum_{\ell} \frac{\ell(\ell+1)}{2\pi}w_{b\ell} N_\ell$.
Since the observing patch and the scan strategy are the same, the patch size $f_\mathrm{sky}$ and the mode mixing matrix $w_{b\ell}$ are the same between \PBXX{} and the new data set. 
Therefore, the ratio of $N_b$ between \PBXX{} and the new one corresponds to the ratio of the total data volume, $n_\mathrm{det} t_\mathrm{obs}$, weighted by $N_{\mathrm{NET}}^{-2}$.
We obtain $+106.5\%$ for the lowest $\ell$ bin and $+95.0\%$ on average.

\subsection{Simulation of the Power Spectrum with Additional Data}
Next, we perform an MC simulation to compute the expected measurements of power spectra by increasing statistics from \PBXX{}.
Here, we do not simulate detector timestreams or maps, but directly simulate power spectra assuming their statistical behavior.
We simulate a set of correlated first and second power spectrum measurements so that their statistical uncertainties are equal to \PBXX{} and the new measurements.
We select simulations whose first measurement is the same as \PBXX{}.
Then, the second measurements of these simulations are the statistical expectations of the new measurement.
The details of the calculation are explained in Appendix~\ref{sec:comparison_appendix}.

\Cref{fig:MCps} shows the results from 100,000 simulations for \textsc{Polarbear} autospectrum and cross spectra with Planck 143, 217, and $353\,\mathrm{GHz}$. 
Here, the MC simulations of the second measurement have shifts from \PBXX{}, even though we require the first measurement to be \PBXX{}.
This is because the second measurement tends to approach the true signal we assume, which is the lensing $B$-mode with Galactic dust foreground and $r=0$.

\subsection{Consistency Probabilities}

\begin{deluxetable}{crrrr}
\tablecaption{Comparison PTE for Each $\ell$-bin\label{tab:compPTE}}
\tabletypesize{\small}
\tablehead{\colhead{} & 
\colhead{$\mathrm{PB}\times\mathrm{PB}$} &
\colhead{$\mathrm{143}\times\mathrm{PB}$} &
\colhead{$\mathrm{217}\times\mathrm{PB}$} &
\colhead{$\mathrm{353}\times\mathrm{PB}$}
}
\startdata
Band definition \\
$50 \le \ell < 100$ & 44.4\% & 26.2\% & 32.2\% & 4.0\% \\
$100 < \ell < 150$ & 21.0\% & 39.2\% & 45.9\% & 93.1\% \\
$150 < \ell < 200$ & 71.7\% & 53.5\% & 67.6\% & 11.2\% \\
$200 < \ell < 250$ & 88.7\% & 14.5\% & 49.2\% & 69.5\% \\
$250 < \ell < 300$ & 9.7\% & 83.5\% & 70.0\% & 53.5\% \\
$300 < \ell < 350$ & 82.4\% & 9.7\% & 2.8\% & 75.2\% \\
$350 < \ell < 400$ & 5.7\% & 6.6\% & 57.6\% & 29.0\% \\
$400 < \ell < 450$ & 45.3\% & 87.5\% & 12.7\% & 97.0\% \\
$450 < \ell < 500$ & 35.4\% & 55.5\% & 33.4\% & 75.9\% \\
$500 < \ell < 550$ & 18.6\% & 61.4\% & 35.6\% & 0.8\% \\
$550 < \ell < 600$ & 75.6\% & 7.9\% & 49.6\% & 0.2\% 
\enddata
\vspace*{-20pt}
\end{deluxetable}
\begin{deluxetable}{crrrr}
\tablecaption{Comparison PTE for Each Spectrum\label{tab:compPTEspec}}
\tabletypesize{\small}
\tablehead{\colhead{} & 
\colhead{$\mathrm{PB}\times\mathrm{PB}$} &
\colhead{$\mathrm{143}\times\mathrm{PB}$} &
\colhead{$\mathrm{217}\times\mathrm{PB}$} &
\colhead{$\mathrm{353}\times\mathrm{PB}$}
}
\startdata
Total $\chi^2$ & 36.4\% & 21.6\% & 37.1\% & 0.9\% \\
(dof: 11) & (12.1) & (14.4) & (11.9) & (25.2) \\\hline KS test & 85.3\% & 61.2\% & 22.7\% & 42.4\% \\
\enddata
\tablecomments{
The number in parentheses shows the total $\chi^2$ value.}
\end{deluxetable}

\vspace*{-2\intextsep}
We compare the new result in this work with these MC simulations and evaluate PTEs for each spectrum and each $\ell$ band.
The results are shown in \Cref{tab:compPTE}.
We find no extremely low PTEs in the \textsc{Polarbear} autospectrum and cross-spectrum with Planck $143\,\mathrm{GHz}$.
In cross spectra with Planck $217$ and $353\,\mathrm{GHz}$, we find some low PTEs.
\Cref{tab:compPTEspec} also shows the total $\chi^2$ of 11 $\ell$-bins and its PTE. We find again that the cross-spectrum with Planck $353\,\mathrm{GHz}$ has a low PTE.
The overall $\chi^2$ of all 44 $\ell$ bins is 63.6, and its PTE is 3.0\%.

We also perform the KS test to compare the distribution of PTEs with a uniform distribution.
The results for each spectrum are shown in \Cref{tab:compPTEspec}.
The overall KS test PTE from all 44 PTEs in \Cref{tab:compPTE} is 44.0\%. In this case, we do not find any low PTEs.

Although the overall $\chi^2$ PTE is low, it is due to the two highest $\ell$ points of the cross-spectrum with Planck $353\,\mathrm{GHz}$.
The positive shift at $550<\ell<600$ may indicate the contamination of polarized dusty star-forming galaxies \citep{Bonavera2017,Lagache2020}, although this would not explain the negative shift at $500<\ell<550$. 
If we remove the highest $\ell$ point, the total $\chi^2$ of the spectrum becomes 15.7, and its PTE improves to 11.0\%. The overall $\chi^2$ PTE also improves to 12.4\%.
The impacts of the two lowest PTE bins on the tensor-to-scalar ratio $r$ are smaller than $2\times10^{-4}$.
Since all PTEs for \textsc{Polarbear} autospectrum are reasonable, 
we conclude that the \PBXX{} result and the new result are consistent in terms of the $r$ measurements. 
Future measurements with more statistics will give a conclusive understanding of the cross-spectrum with Planck $353\,\mathrm{GHz}$.

Finally, we apply this consistency check to the evaluation of the tensor-to-scalar ratio $r$.
Since lower $\ell$ bins are more sensitive to $r$, the overall PTE changes.
We compare the estimate of $r$ from the new measurement with those from the MC simulations.
Note again that MC simulations assume a fiducial true signal with $r=0$.
Only for consistency checking, we use a naive estimation compared to that performed in \cref{sec:pe}.
We fit only $r$ by fixing the foreground parameters.
We obtain the distribution of the best-fit $r$ from MC simulations as $r = 0.08_{-0.12}^{+0.13}$.
Here, the bias comes from the contribution of the first measurement fixed to the result of \PBXX{}.
By comparing the actual best-fit $r=-0.04$ from the new data set, we obtain the PTE of 32.9\%. 

\section{Conclusion\label{sec:conclusion}}
We perform an improved analysis of the \PBXX{} data, the three seasons of \textsc{Polarbear} observations on a $670\,\deg^2$ patch of the sky. We successfully recover 80\% more data by improving the angle error correction of the rotating HWP.

By processing the data using the same analysis pipeline as was used in \PBXX{},
we measure the CMB $B$-mode power spectrum at the multipole range $50<\ell<600$. 
We find no excess signal beyond that expected from the combination of a $\Lambda\mathrm{CDM}$ model and a Galactic dust foreground.
We place an upper limit on the tensor-to-scalar ratio $r < 0.33$ at a 95\% confidence level.
The change in the \textsc{Polarbear} $B$-mode power spectrum from \PBXX{} is consistent with statistical expectations due to the additional data.

Our result demonstrates the possibility of measuring degree-scale CMB anisotropies with a ground-based telescope located in the Atacama desert of Chile.
The rotating HWP is a key technique for separating contamination from atmospheric fluctuations and achieving a good noise performance at low frequencies. 
The low-frequency noise performance enabled by the continuously rotating HWP is one of reasons why some future experiments, including the Simons Array experiment~\citep{SA2016JLTP} and the Small Aperture Telescope of the Simons Observatory experiment~\citep{SO2019JCAP}, plan to employ rotating HWPs.
As we show in this paper, accurate angle encoding is a key to the success of this methodology. \vfill~

\begin{acknowledgments}
The \textsc{Polarbear} project is funded by the National Science Foundation under grants AST-0618398 and AST-1212230.
The analysis presented here was also supported by Moore Foundation grant number 4633, the Simons Foundation grant number 034079, and the Templeton Foundation grant number 58724.
The James Ax Observatory operates in the Parque Astronómico Atacama in Northern Chile under the auspices of the Comisión Nacional de Investigación Científica y Tecnológica de Chile (CONICYT).
Work at LBNL is supported in part by the U.S.\ Department of Energy, Office of Science, Office of High Energy Physics, under contract No.\ DE-AC02-05CH11231.
This research used resources of the National Energy Research Scientific Computing Center, which is supported by the Office of Science of the U.S.\ Department of Energy under Contract No.\ DE-AC02-05CH11231.
In Japan, this work was supported by MEXT KAKENHI Grant Numbers JP21111002, JP15H05891, JP18H05539, and JP18H04362, JSPS KAKENHI Grant Numbers JP26220709, JP26800125, JP14J01662, JP16K21744, JP17K18785, JP18H01240, JP18J02133, JP18K13558, JP19H00674, JP19KK0079, JP20H01921, JP20K14481, JP20K14497, JP21K03585, and the JSPS Core-to-Core Program.
This research used resources of the Central Computing System, owned and operated by the Computing Research Center at KEK.
This work was supported by the World Premier International Research Center Initiative (WPI), MEXT, Japan.
Support from the Ax Center for Experimental Cosmology at UC San Diego is gratefully acknowledged.
The SISSA group acknowledges the INDARK INFN Initiative and the COSMOS and LiteBIRD networks of the Italian Space Agency (cosmosnet.it).
The Melbourne group acknowledges support from the Australian Research Council’s Discovery Projects scheme (DP210102386). 
The Dalhousie group acknowledges support from the Natural Sciences and Engineering Research Council (NSERC) and Canada Foundation for Innovation (CFI).
M.A.\ acknowledges support from CONICYT UC Berkeley-Chile Seed Grant (CLAS fund) Number 77047, Fondecyt project 1130777 and 1171811, DFI postgraduate scholarship program and DFI Postgraduate Competitive Fund for Support in the Attendance to Scientific Events.
G.F.\ acknowledges the support of the European Research Council under the Marie Sk\l{}odowska Curie actions through the Individual Global Fellowship No.~892401 PiCOGAMBAS.
D.P.\ acknowledges support from the COSMOS Network from the Italian Space Agency.

\vfill~
\end{acknowledgments}

\appendix
\section{Simulation of power spectrum measurements with additional data} \label{sec:comparison_appendix}
Here, we describe details of the MC simulation to generate first and second power spectrum measurements that are correlated due to the common data. 

Consider a case that the observed \textsc{Polarbear} autospectrum $\hat{D}_b$ and noise spectrum $N_b$ in the first measurement are updated to $\hat{D}'_b$ and $N'_b$ in the new measurement with additional data.
As explained in \cref{sec:effadd}, the noise spectrum is proportional to the inverse of the data volume and determined as $N'_b = N_b / (1+\alpha_b)$, where $\alpha_b$ is the effective fractional increase of data for each $\ell$ band.

On the other hand, the observed spectra $\hat{D}_b$ and $\hat{D}'_b$ are random variables dependent on realizations of noise.
We approximate that each bin of the noise bandpower spectrum follows a $\chi$-squared distribution with $\nu_b$ degrees of freedom and the mean value of $N_b$.
We can make one realization of the first measurement following the distribution as
\begin{equation}
\hat{D}_b + N_b = D_b + \frac{N_b}{\nu_b} {\sum_{i=1}^{\nu_b} X_{b,i}^2}\;,\label{eq:genMC1}
\end{equation}
where $D_b$ is the true signal, and $X_{b,i}$ is each of $\nu_b$ independent random numbers that follow a normal distribution with the variance $\langle X_{b,i}^2 \rangle = 1$.
The physical meaning of $X_{b,i}$ is a measurement error of the amplitude of each mode of anisotropies, $a_{\ell m}$, normalized by the noise spectrum.
We ignore the covariance between the true signal and noise for simplicity.

By increasing the data volume, we can improve the measurement of $X_{b,i}$ as
\begin{equation}
X'_{b,i} = (X_{b,i} + \alpha Y_{b,i}) / (1 + \alpha_b)\;,
\end{equation}
where $Y_{b,i}$ is another set of random numbers that follow a normal distribution.
Its variance is $\langle Y_{b,i}^2 \rangle=1/\alpha_b$ because the volume of additional data is $\alpha_b$ times as large as the first volume.
The realization of the new measurement becomes
\begin{equation}
\hat{D}'_b + N'_b = D_b + \frac{N_b}{\nu_b} {\sum_{i=1}^{\nu_b} X^{\prime 2}_{b,i}}\;. \label{eq:genMC2}
\end{equation}
By taking the average of \cref{eq:genMC2} of the MC realizations, we obtain $N'_b = N_b / (1+\alpha_b)$, which agrees with the above discussion about the data volume.

The first measurement $\hat{D}_b$ obtained above, however, is random. In order to evaluate the conditional probability having the first measurement as \PBXX{}, we repeat a trial of drawing a set of $X_{b,i}$ until $\hat{D}_b$ becomes sufficiently close to the \PBXX{} result. 
On the other hand, we do not apply any condition on $Y_{b,i}$ and the resulting $\hat{D}'_b$.
To reduce the computational cost, we reuse the same $X_{b,i}$ for 100 draws of $Y_{b,i}$.
The total number of MC realizations is 100,000 with 1000 draws of $X_{b,i}$.

Next, we extend this method to the cross-spectrum with Planck.
Although the Planck data are the same for the first and second measurements, the change of \textsc{Polarbear} data affects the cross-spectrum.
Similar to \cref{eq:genMC1,eq:genMC2}, the cross spectra with one of Planck frequencies for the first and second measurements are computed as
\begin{equation}
\hat{D}_{b,\mathrm{Planck}\times\mathrm{PB}} = D_{b,\mathrm{Planck}\times\mathrm{PB}} + \frac{\sqrt{N_b^{\mathrm{Planck}} N_b}}{\nu_b}\sum_{i=1}^{\nu_b}Z_{b,i}X_{b,i}\;,
\end{equation}
and
\begin{equation}
\hat{D}_{b,\mathrm{Planck}\times\mathrm{PB}}^{\prime} = D_{b,\mathrm{Planck}\times\mathrm{PB}} + \frac{\sqrt{N_b^{\mathrm{Planck}}N_b}}{\nu_b}\sum_{i=1}^{\nu_b}Z_{b,i}X'_{b,i}\;,
\end{equation}
respectively, where $D_{b,\mathrm{Planck}\times\mathrm{PB}}$ is the assumed signal spectrum depending on the Planck frequency, $N_b^{\mathrm{Planck}}$ is the noise spectrum for the Planck frequency, and $Z_{b,i}$ is a set of random numbers that follows a normal distribution with a variance of 1. The noise bias in the cross-spectrum becomes zero.

To apply the requirement on the first measurement, 
after finding a set of $X_{b,i}$ as described above,
we repeat drawing a set of $Z_{b,i}$ until $\hat{D}_{b,\mathrm{Planck}\times\mathrm{PB}}$ becomes sufficiently close to the result of \PBXX{}.
We use the same set of $X_{b,i}$ and $Z_{b,i}$ for 100 draws of $Y_{b,i}$.

We do not simulate Planck auto- and cross spectra among Planck frequencies.
We use the same input Planck maps for both \PBXX{} and this analysis.
Since we scan the Planck map using the \textsc{Polarbear} observation, 
the data increase slightly modifies the shape of the mask used for truncation.
We find that the impact is small enough, however, and the spectra are almost the same.
\clearpage

\bibliography{reference}{}
\bibliographystyle{aasjournal}

\allauthors

\end{document}